\documentclass[apj]{emulateapj}
\usepackage{apjfonts}
\usepackage{amsbsy}

\newcommand{\up}[1]{{\rm #1}}
\newcommand{\plin}{P_\up{lin}}
\newcommand{\pR}{P_\up{R}}
\newcommand{\wpp}{w_p(r_p)}

\newcommand{\kpc}{\ {\rm kpc}}
\newcommand{\mpc}{{\rm Mpc}}

\newcommand{\hkpc}{{h^{-1}\kpc}}
\newcommand{\hmpc}{{h^{-1}\mpc}}

\newcommand{\hmpci}{{h\mpc^{-1}}}
\newcommand{\OM}{\Omega_m}
\newcommand{\OB}{\Omega_b}
\newcommand{\OD}{\Omega_\Lambda}

\newcommand{\rms}{\sigma_8}
\newcommand{\Mmin}{M_\up{min}}
\newcommand{\Mcut}{M_\up{cut}}

\newcommand{\Rvir}{R_\up{vir}}

\newcommand{\sigh}{\sigma_\up{h}}
\newcommand{\Nsat}{N_\up{sat}}

\newcommand{\NsatA}{\langle\Nsat\rangle}

\newcommand{\vv}{\upsilon}
\newcommand{\xiR}{\xi_\up{R}}
\newcommand{\xipp}{\xi(r_\parallel,r_\perp)}
\newcommand{\kms}{{\rm km\, s}^{-1}}
\newcommand{\ppp}{P(k_\parallel,k_\perp)}
\newcommand{\pmm}{P_{mm}}
\newcommand{\ximm}{\xi_{mm}}
\newcommand{\msun}{M_{\odot}}
\newcommand{\hmsun}{{h^{-1}\msun}}
\newcommand{\dimp}{\Delta^2(k)}
\newcommand{\pnl}{P_\up{nl}}

\newcommand{\pzr}{P_\up{Z\rightarrow R}}
\newcommand{\pzrF}[1]{P^{#1}_\up{Z\rightarrow R}}

\newcommand{\rpp}{r_p}
\newcommand{\pobs}{P_\up{obs}}

\newcommand{\beeq}{\vspace{10pt}\begin{equation}}
\newcommand{\eneq}{\vspace{10pt}\end{equation}}

\begin{document}

\title{Extending Recovery of the Primordial Matter Power Spectrum}

\author{Jaiyul Yoo$^{1,2,3}$, David H. Weinberg$^3$, 
Jeremy L. Tinker$^4$, Zheng Zheng$^{5,6}$, and Michael S. Warren$^7$}
\altaffiltext{1}{Harvard-Smithsonian Center for Astrophysics,
Harvard University, 60~Garden Street, Cambridge, MA 02138; jyoo@cfa.harvard.edu}
\altaffiltext{2}{Menzel Fellow}
\altaffiltext{3}{Department of Astronomy, The Ohio State University, 
140 West, 18th Avenue, Columbus, OH 43210; dhw@astronomy.ohio-state.edu}
\altaffiltext{4}{Kavli Institute for Cosmological Physics, 
University of Chicago, Chicago, IL 60637; tinker@cfcp.uchicago.edu}
\altaffiltext{5}{School of Natural Sciences, Institute for Advanced Study,
Einstein Drive, Princeton, NJ 08540; zhengz@ias.edu}
\altaffiltext{6}{John Bahcall Fellow}
\altaffiltext{7}{Theoretical Astrophysics Division, Los Alamos National 
Laboratory, Los Alamos, NM 87543; msw@lanl.gov}

\slugcomment{submitted to The Astrophysical Journal}
\shorttitle{PRIMORDIAL MATTER POWER SPECTRUM RECOVERY}
\shortauthors{YOO ET AL.}

\begin{abstract}
The shape of the primordial matter power spectrum encodes critical
information on cosmological parameters. At large scales, in the linear regime,
the observable galaxy power spectrum $\pobs(k)$ is expected to follow the 
shape of the linear matter power spectrum $\plin(k)$,
but on smaller scales the effects
of nonlinearity and galaxy bias make the ratio $\pobs(k)/\plin(k)$ 
scale-dependent.
We develop a method that can extend the dynamic range of the primordial matter
power spectrum recovery, taking full advantage of precision measurements on
quasi-linear scales, by incorporating additional constraints on the galaxy
halo occupation distribution (HOD) from the projected galaxy correlation 
function $\wpp$. We devise an analytic model to calculate
observable galaxy power spectrum $\pobs(k)$ in real-space and
redshift-space, given $\plin(k)$ and HOD parameters, and we demonstrate its
accuracy at the few percent level with tests
against a suite of populated $N$-body simulations. Once HOD parameters are 
determined by fitting $\wpp$ measurements for a given cosmological model, 
galaxy bias is completely specified, and our analytic model predicts both the
shape and normalization of $\pobs(k)$. Applying our method to 
the main galaxy redshift samples from the Sloan Digital Sky Survey (SDSS), 
we find that the real-space galaxy power spectrum follows the shape of the
nonlinear matter power spectrum at the $1-2\%$ level up to $k=0.2\hmpci$ and 
that current observational uncertainties in HOD parameters leave only few 
percent uncertainties in our scale-dependent bias predictions up to 
$k=0.5\hmpci$. These uncertainties can be marginalized over in deriving 
cosmological parameter constraints, and they can be reduced by higher 
precision $\wpp$ measurements. When we apply our method to the SDSS luminous 
red galaxy (LRG) samples, we find that the
linear bias approximation is accurate to 5\% at $k\leq0.08\hmpci$, but the 
strong scale-dependence of LRG bias prevents the use of linear theory at
$k\geq0.08\hmpci$. Our HOD model prediction is in good agreement with the 
recent SDSS LRG power spectrum measurements at all measured
scales ($k\leq0.2\hmpci$), naturally explaining the observed shape of 
$\pobs(k)$ in the quasi-linear regime. The phenomenological ''$Q$-model''
prescription is a poor description of
galaxy bias for the LRG samples, and it can lead to biased cosmological
parameter estimates when measurements at $k\geq0.1\hmpci$ are included in the 
analysis. We quantify the potential bias and constraints on cosmological 
parameters that arise from applying linear theory and $Q$-model fitting, 
and we demonstrate the utility of HOD modeling of high precision measurements 
of $\pobs(k)$ on quasi-linear scales, which will be obtainable from 
the final SDSS data set.
\end{abstract}


\section{Introduction}
\label{sec:int}
In the linear regime, the power spectrum of matter fluctuations encodes
information about the physics of early universe (e.g., the potential of the
field that drives inflation) and about the matter and energy contents of 
the cosmos. The power
spectrum of galaxies can be biased relative to the power spectrum of matter
\citep{nick,BBKS}, but fairly general theoretical arguments imply that 
the shape of
galaxy power spectrum should approach the shape of the linear matter power 
spectrum $\plin(k)$ at sufficiently large scales, i.e., 
\beeq
\pR(k)=b_0^2~\plin(k)+N_0,
\label{eq:bias0}
\eneq
where $b_0$ is a constant galaxy bias factor and 
$\pR(k)$ denotes the real-space galaxy power spectrum 
\citep{peter,fry,dhw95,mann,bob2,vijay,andreas,alexia}.
The additive ``shot noise'' term $N_0$ reflects both galaxy discreteness and
small scale clustering \citep{bob2,patMc,smith2006}; in general, it can
differ from a simple Poisson sampling correction.
In the linear regime, distortions of redshift-space structure by peculiar 
velocities also alter the amplitude but not the shape of galaxy power spectrum 
\citep{nick2}.
There have therefore been great efforts to measure the galaxy power spectrum on
large scales from angular catalogs and redshift surveys and 
to use the results to
test cosmological models (e.g., \citealt{yupee,baum,feldman,cbp,lcrs,pscrs}).
The enormous size of the Two Degree Field Galaxy Redshift Survey (2dFGRS;
\citealt{colless}) and the Sloan Digital Sky Survey (SDSS;
\citealt{sloan}) relative to earlier samples allows much higher precision
measurements. The current state-of-the-art power spectrum measurements are
Cole~et~al.'s (2005) 
analysis of the power spectrum in the 2dFGRS, \citet{nikhil}, and 
Blake~et~al.'s (2007)
analyses of luminous red galaxies (LRGs) with photometric redshifts in the
SDSS, and \citet{will2} and Tegmark~et~al.'s (2006)
measurements from the SDSS redshift survey of
main sample galaxies and LRGs.

This paper investigates the problem of going from the galaxy power spectrum to
the linear matter power spectrum, and hence to cosmological conclusions.
The latest observational analyses yield 
impressive statistical precision on scales near the
transition from the
linear to the nonlinear regime, e.g., typical 1-$\sigma$ errors
of 5$-$10\% in $P(k)$ at $k\simeq0.15\hmpci$. The critical uncertainty in 
cosmological
interpretation is therefore the accuracy of equation~(\ref{eq:bias0}) on these
scales. The effects of nonlinearity and redshift-space distortions on the
$matter$
power spectrum can be computed using numerical simulations or tuned analytic
models \citep[and references therein]{smith},
but details of galaxy formation physics can influence the relation
between galaxy and matter power spectra in this regime. \citet{will2} 
find that linear theory fits
imply different cosmological parameters if applied to 
measurements with $k\leq0.06\hmpci$ or 
with $k\leq0.15\hmpci$, indicating that nonlinear effects have become
significant in this regime. Furthermore, \citet{asp} 
analyze the SDSS and 2dFGRS galaxy samples 
and find that the measured shapes of galaxy power spectra differ
at a level that cannot be explained by the
expected cosmic variance. They show that the
likely source of the discrepancy is different scale-dependence of galaxy bias,
originating from the different color distributions of galaxies
in the SDSS and 2dFGRS samples.

\citet{2dfn}, \citet{max2} and \citet{nikhil}
approach this problem by fitting a parametrized model of
scale-dependent bias,
\beeq
P_\up{gal}(k)=b_0^2\plin(k){1+Qk^2\over1+Ak},
\label{eq:qmodel}
\eneq
where we use $P_\up{gal}(k)$ to represent the galaxy power spectrum,
which can be either in real-space or redshift-space.
The functional form is devised for convenience
to approximate the scale-dependent bias of galaxy samples obtained by 
populating the Hubble volume simulation \citep{hubblevol} using a 
semi-analytic model of galaxy formation \citep{andrew1}.
Here $A=1.4\hmpc$ or $1.7\hmpc$ for real-space power spectrum $\pR(k)$ or 
angle-averaged
redshift-space power spectrum $P_0(k)$ measurements, respectively, and $Q$ is
treated as a free parameter that is marginalized over in deriving cosmological
parameter constraints. This approach is adequate {\it if}
equation~(\ref{eq:qmodel}) is a sufficiently accurate description of 
scale-dependent bias for some value of $Q$, but it could yield biased 
parameter estimates or incorrect error bars if the actual scale-dependence
is different. It also gives up on extracting cosmological information from
scales where bias might be mildly scale-dependent. For example, 
\citet[hereafter, T06]{max2}
find that cosmological parameters remain unaffected by changes in power
spectrum measurements at $k\geq0.1\hmpci$ 
once they marginalize over the value of $Q$.
This implies that the statistical constraining power on cosmological parameters
is lost at $k\geq0.1\hmpci$ by the marginalization process.

In this paper, we present an alternative approach to recovering the
shape of the linear
matter power spectrum, both more aggressive and more robust than
``marginalizing over $Q$.'' Our approach is based
on the halo occupation distribution (HOD) framework, which describes the
nonlinear relation between galaxies and matter by specifying the probability
$P(N|M)$ that a halo of mass $M$ hosts $N$ number of galaxies of a given
type, together with specification of the relative spatial and velocity 
distributions of galaxies within halos.\footnote{Throughout this paper, 
the term ``halo'' refers to a
dark matter structure of overdensity $\rho/\bar \rho_m\simeq200$,
in approximate dynamical equilibrium.}
The HOD formalism has emerged as a powerful method of modeling galaxy bias
\citep{jing2,uros,chung,john2,roman,andreas} because the dynamics of dark matter
halos can be accurately calculated using analytic approximations or
$N$-body simulations, and the effects of galaxy 
formation physics can be parametrized in terms of an HOD and inferred by
fitting observational data.

Our strategy for extending recovery of the primordial matter power spectrum
is to use complementary information from the measurements of the
projected correlation function $\wpp$ as a constraint to obtain HOD parameters
given a cosmological model. We then predict the galaxy power spectrum 
$P_\up{gal}(k)$ and study the scale-dependent bias
\beeq
b^2(k)\equiv P_\up{gal}(k)/\plin(k).
\label{eq:scale}
\eneq
For each cosmological model, fitting $\wpp$ measurements determines HOD 
parameters and we can then compute a {\it unique} prediction of $P_\up{gal}(k)$,
both shape and normalization (which is essentially pinned to the amplitude
of $\wpp$).
Uncertainties in HOD parameters introduce uncertainty in $P_\up{gal}(k)$ and
$b^2(k)$,
but these uncertainties can be accurately computed and marginalized over. 
Therefore, we can extend the wavenumber range over which
$P_\up{gal}(k)$ measurements can be used for cosmological parameter 
constraints, 
taking full advantage of precision measurements on quasi-linear scales.
In practice, we are just using the measured $P_\up{gal}(k)$ 
and $\wpp$ to simultaneously constrain
HOD parameters and the cosmological parameters and marginalizing over
the former. Relative to the
$Q$-model approach, our method adopts a more physically motivated computation
of
$P_\up{gal}(k)$ and $b^2(k)$, requiring only the validity of the adopted HOD 
parametrization, and it brings in the additional information present in $\wpp$
rather than using only the $P(k)$ shape itself to constrain the 
scale-dependence of bias.

In principle, the power spectrum $P(k)$ 
and correlation function $\xi(r)$ contain
the same information. However, they are in practice measured via different
estimators and on different scales, where their signal-to-noise ratios are 
highest and systematic errors are 
relatively well
understood. The information in $P(k)$ and $\xi(r)$ measurements on these
non-overlapping scales is therefore not identical, but complementary.
Furthermore, the projected correlation function $\wpp$ is measured
to ease the difficulty in interpreting nonlinear
redshift-space distortion of correlation function measurements
on small scales. Therefore, the addition of $\wpp$ measurements 
at $r_p\leq30\hmpc$ brings new information that is not present in $P(k)$ 
measurements at $k\leq1\hmpci$.

A different
approach to this problem is to develop an analytic model for predicting the
scale-dependence of galaxy bias by using higher-order perturbation theory
(e.g., \citealt{patMc,smith2006}). This approach is elegant and transparent
in nature, since it is based on linear theory and its extension to 
higher-order, while our approach is less {\it ab initio} in the sense of
incorporating elements calibrated by numerical $N$-body simulations in our
analytic model. However, the critical uncertainty for this approach based
on higher-order perturbation theory is its applicability on quasi-linear
scales ($\gtrsim0.1\hmpci$), where first-order linear theory is known to be
inaccurate, but the measurement precision is highest in practice. In contrast,
our approach is fully nonlinear, and phenomenological in nature, 
so it can be applied down to small scales, limited only by the point at which
uncertainties in the HOD parameters introduce systematic uncertainty in the
$P(k)$ recovery.

Analyses of galaxy redshift surveys typically estimate 
the angle-averaged power spectrum
$P_0(k)$, i.e., the monopole of the redshift-space power spectrum (e.g., 
\citealt{2dfn,will2}). Redshift-space distortions do not alter the shape in 
linear theory, but they do change the shape in the trans-linear regime (e.g., 
\citealt{cole1}), and finger-of-god (FoG) effects have impact out to 
large scales (e.g., \citealt{roman}). \citet{nikhil} and \citet{blake}
deproject the angular clustering measurements of the SDSS LRG sample using
photo-$z$ catalogs to estimate the real-space power spectrum, independent of 
redshift-space distortions. \citet{max1,max2} use a linear 
combination of the redshift-space monopole, quadrupole, and hexadecapole that
recovers the real-space power spectrum in the linear regime. We will denote
this ``pseudo real-space'' power spectrum $\pzr(k)$. The
redshift-space power spectrum
estimators can be applied directly to galaxy redshift data or 
applied after compressing FoG effects.
We will investigate $P_\up{gal}(k)$ and $b^2(k)$ for all of these cases.

To this end, we develop an analytic model in \S~\ref{sec:method}
for calculating
real-space and redshift-space galaxy power spectra given $\plin(k)$ and a
galaxy HOD, drawing on the \citet{jeremy2} model for redshift-space distortion,
which improves on previous work (e.g., \citealt{uros3,martin,kang,coor}). 
\citet{jeremy2} tests the model for computing the
redshift-space correlation function against
a series of populated $N$-body simulations. Here we extend the model and 
present additional tests of its applicability to modeling redshift-space power 
spectra in \S~\ref{sec:gal}.

In this paper, we use HOD parameters for volume-limited galaxy samples that
have well defined classes of galaxies, focusing on SDSS main galaxy samples
with absolute-magnitude limits $M_r\leq-20$ and $M_r\leq-21$ \citep{idit3}
in \S~\ref{sec:lin}, and SDSS LRG samples with absolute-magnitude limits
$-23.2\leq M_g\leq-21.2$ and $-23.2\leq M_g\leq-21.8$ 
\citep{daniel,idit4,zheng5} in
\S~\ref{sec:lrg} for application of our method.\footnote{For brevity, 
we quote the absolute magnitude thresholds $M_r-5\log h$ and
$M_g-5\log h$ for $h\equiv1$.} More
complete modeling of the conditional luminosity function \citep{yang}
might allow use of 
flux-limited galaxy catalogs, though it requires more free parameters to
provide complete descriptions of the galaxy samples. Here we only consider
volume-limited galaxy samples, whose results can be combined to improve 
statistical precision. We summarize our main results in \S~\ref{sec:sum}.

\section{Calculational Methods}
\label{sec:method}
\subsection{Numerical Model}
\label{sec:numerical}
We use the $N$-body simulations of \citet{jeremy}
to test our analytic model calculations of the
correlation function and the power spectrum in
real-space and redshift-space. These are five simulations of a flat 
$\Lambda$CDM universe using the publicly available tree-code 
{\scriptsize GADGET} \citep{gadget}, and all the
simulations are performed with identical cosmological parameters except for
the random seed numbers used to generate initial conditions.
The initial scale-invariant ($n_s=1$) power spectrum is modified by the 
transfer function of \citet{george} with shape parameter $\Gamma=0.2$. 
The simulation was evolved from an expansion factor $a=0.01$ to $a=1.0$
with $\OM=0.1$, $\OD=0.9$ and $\rms=0.95$ at $z=0$.
To cover a range of parameter space spanned by $\OM$ and $\rms$, 
we use earlier outputs to represent different cosmological 
models from the simulations. Our choices for the earlier 
expansion factors are $a_\up{out}=$0.84, 0.64, 0.49, and 0.40. These outputs
correspond respectively to simulations with different 
parameter combinations ($\OM$, $\rms$)=(0.16, 0.90), (0.30, 0.80), 
(0.48, 0.69), and (0.63, 0.60) with the identical power spectrum shape
($\Gamma=0.2$) but evolved beginning
at expansion factor $a=0.01/a_\up{out}$. 
This procedure correctly provides the density field that would be obtained
from an independent simulation evolved to $z=0$ with
the corresponding parameter 
combination of $\OM$ and $\rms$ \citep{zheng1}.
We evolve 360$^3$ particles in a volume of comoving side length 253~$\hmpc$ 
to take into consideration that the lowest mass halos that host galaxies with
$M_r\leq-20$ contain at least 32 particles. Dark matter halos are identified
by using the friends-of-friends algorithm (FoF; \citealt{fof}) with a linking
length of 0.2 times the mean interparticle separation, i.e., $140\hkpc$.

To populate dark matter halos with galaxies in $N$-body simulations,
we use HOD parameters listed in Table~\ref{tab:par}
that are chosen to match the mean number density $\bar n_g$
and projected correlation functions $\wpp$ of the 
SDSS galaxy samples with absolute-magnitude limits
$M_r\leq-20$ and $M_r\leq-21$ \citep{idit3}. In our standard HOD
parametrization, the number of central galaxies is a step
function changing from zero to one at a minimum halo mass $\Mmin$.
Therefore, halos of mass $M<\Mmin$ lack 
galaxies. We assume $\langle N_\up{sat}\rangle\propto M$ at high masses with a
smooth cutoff at low mass. Therefore, the number of satellite galaxies is, 
\beeq
\langle N_\up{sat}\rangle_M=\left({M\over M_1}\right)\exp\left(-{\Mcut\over M-
\Mmin}\right),
\eneq
for a halo of mass $M\geq\Mmin$, and the distribution of satellite galaxy
number $P(\Nsat|\NsatA)$ is assumed to be Poisson \citep{andrey,zheng3}.
This parametrization is well suited to our purposes, but we also investigate
the effect of adopting a more flexible HOD parametrization
in \S~\ref{sec:lin}.

We replace halos identified by the FoF algorithm by spherical NFW halos 
\citep{nfw} with identical mass, truncated at virial radius $\Rvir$, within
which the mean density is 200 times the mean matter density.
The concentration parameters $c_\up{dm}$ of dark matter halos are computed
using the relation of \citet{james} and are scaled to account for the 
different definition of halo overdensity adopted here.
This NFW-replacement
method reduces numerical artifacts caused by finite force resolution
in our simulations.
We place a central galaxy at the center of mass of halos. 
Assuming that satellite galaxies trace the dark matter distribution 
within halos, we place satellite galaxies following the NFW profile of halos.

In redshift-space, galaxies are displaced because of peculiar velocity.
Central galaxies are assumed to be at rest relative
to the halo center; no velocity bias is assumed for central galaxies. 
For satellite galaxies, we add line-of-sight velocities drawn from
a Gaussian distribution with zero mean and dispersion
\beeq
\sigma_\vv(M)=\left({GM\over2\Rvir}\right)^{1/2}
\label{eq:disp}
\eneq
to the velocity of the halo center of mass. This procedure is
exact for isotropic singular isothermal halos and is reasonably accurate
for NFW profiles (see \citealt{jeremy} for detailed tests).

Finally, we compute the density 
contrast field by cloud-in-cell weighting the particle
distribution onto 360$^3$ grids and use the publicly available fast 
Fourier transform (FFT) code, {\scriptsize FFTW}, 
to obtain the Fourier components in units
of the fundamental mode of our simulation box, $\Delta k=0.02\hmpci$. 
We deconvolve
the cloud-in-cell weighting function $W_\up{CIC}(k)$,
and subtract shot-noise contributions $1/N_\up{gal}$
to compute the power spectrum at each $k$:
\beeq
P_\up{gal}(k)=P_\up{FFT}(k)/W^2_\up{CIC}(k)-{1\over N_\up{gal}},
\eneq
where the weighting function is
\beeq
W_\up{CIC}(k)=\left[\prod_{i=1}^{3}
{\sin(\pi k_i/2k_N)\over\pi k_i/2k_N}\right]^2
\eneq
and $N_\up{gal}$ is the number of galaxies, $k_i$ is the $i$-th component
of wavenumber $k$, and
$k_N=4.5\hmpci$ is the Nyquist wavenumber of our simulations.
For computations in redshift-space,
we simply displace particles using the $z$-component of the peculiar velocity
scaled by the Hubble constant
as they would appear to a distant observer at $z=-\infty$.
This procedure satisfies the distant observer approximation we adopt
here, and it ensures periodic 
radial velocity fields in the simulation volume, appropriate 
for FFT. Redshift-space multipoles are extracted by least-squares fitting to
the Legendre polynomial coefficients (eq.[\ref{eq:mulp}]).
We repeat the procedure for the
$x$- and $y$- axes, treating each axis as the line-of-sight, and we average the
resulting power spectra over the three line-of-sight directions.

\subsection{Analytic Model}
\label{sec:analytic}
Our analytic calculation of the real-space galaxy auto-correlation function 
$\xiR(r)$ follows \citet{mtl}, which improves the method of \citet{zheng2}
with more accurate
treatments of scale-dependent halo bias and halo exclusion.
For a given galaxy sample with its
projected correlation function measurements $\wpp$, we obtain HOD parameters
by fitting the mean space density $\bar n_g$ and $\wpp$, computed by
\beeq
\bar n_g=\int_{\Mmin}^\infty dM {dn\over dM}\langle N\rangle_M,
\eneq
\beeq
\wpp=2\int_0^{z_\up{max}} dz~\xiR\left[(\rpp^2+z^2)^{1/2}\right],
\eneq
where $dn/dM$ is the halo mass function of \citet{jenkins} and
we use $z_\up{max}=40\hmpc$ as adopted in SDSS clustering measurements 
(e.g., \citealt{idit2,idit4,idit3}).
The real-space galaxy power spectrum $\pR(k)$ is computed by taking the Fourier
transform of the correlation $\xiR(r)$.\footnote{In general
the correlation $\xiR^\up{2h}(r)$ from galaxy pairs in two distinct halos is
computed already in Fourier space, and one only needs to add the Fourier
transform of the correlation $\xiR^\up{1h}(r)$ from galaxy pairs in the
same halo to compute $\pR(k)$. However, $\xiR^\up{2h}(r)$ is modified to
account for the finite extent of halos and the scale-dependence of halo bias,
and hence the contributions to $\pR(k)$ from pairs in two distinct halos cannot
be expressed in a simple form in Fourier space
(for details, see \citealt{zheng2,mtl,lensing}).}

We compute the redshift-space correlation function 
$\xipp$ using the probability
distribution of galaxy pairwise velocities $f(\vv_z,r)$,
\beeq
1+\xipp=\int_{-\infty}^\infty d\vv_z\left[1+\xi(r)\right]f(\vv_z,r),
\label{eq:stream}
\eneq
where $r_\perp$ is the projected separation,
$r_\parallel$ is the line-of-sight separation in redshift-space, and 
$\vv_z=100~\kms(r_\parallel-z)/\hmpc$ is the pairwise velocity of galaxies
separated by $r=(r_\perp^2+z^2)^{1/2}$.
Equation~(\ref{eq:stream}) is called the streaming model 
and has been used to measure the mean matter
density \citep{john3,2dfn} and to model redshift-space correlations
\citep{martin,uros3}. 
It is valid in the linear and nonlinear regime \citep{karl,roman2}
provided that the correct $f(\vv_z,r)$ is used.
We adopt the probability
distribution function of \citet{jeremy2} for the galaxy pairwise velocities,
which is an analytic model with some elements calibrated on
$N$-body simulations. We refer the reader to the work by \citet{jeremy2}
for extensive discussion and tests.

\subsection{Redshift-Space Multipoles}
\label{sec:redmult}
Using the analytic model, we
compute the redshift-space correlation function $\xipp$ given $\plin(k)$ and a
set of HOD 
parameters, and we expand $\xipp$ with Legendre polynomials, 
\beeq
\xi(r,\mu)=\sum_{l=0}^{\infty} L_l(\mu)\xi_l(r),
\label{eq:xipp}
\eneq
where $r=(r_\perp^2+r_\parallel^2)^{1/2}$ and $\mu=r_\parallel/r$ is the 
direction cosine of the separation and line-of-sight vectors. 
The redshift-space multipole component is then
\beeq
\xi_l(r)={2l+1\over2}\int_{-1}^{1}d\mu L_l(\mu)\xi(r,\mu).
\eneq
We use $L_l(\mu)$ to denote Legendre polynomials to avoid confusion
with redshift-space multipole power spectra $P_l(k)$ defined below.
The redshift-space power spectrum $\ppp$ can be similarly expanded 
using Legendre polynomials,
\beeq
P(k,\mu)=\sum_{l=0}^{\infty} L_l(\mu)P_l(k),
\label{eq:mulp}
\eneq
where $k=(k_\perp^2+k_\parallel^2)^{1/2}$ and $\mu=k_\parallel/k$, in analogy
to quantities in configuration space. The reflection symmetry
of the correlation function and
power spectrum ensures that multipoles with odd $l$ vanish on average.
Making use of the fact that $\xipp$ and
$\ppp$ are Fourier counterparts, each redshift-space 
multipole component is computed by
\beeq
P_l(k)=4\pi i^l\int_0^\infty r^2\xi_l(r)j_l(kr)dr,
\label{eq:ft}
\eneq
where $j_l(x)$ are spherical Bessel functions \citep{cole1,hamilton1}.
Note that the quadrupole ($l=2$) components of $P_l(k)$ and $\xi_l(r)$ have
opposite sign.

Equation~(\ref{eq:ft}) requires knowledge of $\xi_l(r)$ on large
scales, while our analytic model is only tested at $r\leq40\hmpc$.
Therefore, we compute $\xi_l(r)$ at $r\geq40\hmpc$ with the linear
approximation for redshift-space distortion.
In the linear regime, the multipole expansion of $\xipp$
has only three nonzero 
multipoles: monopole $\xi_0$, quadrupole $\xi_2$, and hexadecapole $\xi_4$ 
\citep{nick2}, which are in turn related to $\xiR(r)$,
\begin{eqnarray}
\xi_0(r)&&=C_0\xi_\up{R}(r), \nonumber \\
\xi_2(r)&&=C_2\left(\xiR(r)-\bar\xi(r)\right), \nonumber \\
\xi_4(r)&&=C_4\left(\xiR(r)+2.5\bar\xi(r)-3.5\overline{\overline{\xi}}(r)
\right),
\label{eq:multipole}
\end{eqnarray}
where $C_0=1+{2\over3}\beta+{1\over5}\beta^2$,
$C_2={4\over3}\beta+{4\over7}\beta^2$, $C_4={8\over35}\beta^2$, 
$\beta=\OM^{0.6}/b_0$, $b_0$ is the asymptotic galaxy bias factor,
and the barred correlations are
\beeq
\bar\xi(r)={3\over r^3}\int_0^r s^2\xiR(s)ds,
\eneq
\beeq
\overline{\overline{\xi}}(r)={5\over r^5}\int_0^r s^4\xiR(s)ds
\eneq
\citep{hamilton0}. 
We compute the three multipoles at $r\geq40\hmpc$, rather than
$\xipp$ itself. We first compute the values of $C_l$ at 
$r=40\hmpc$, where the deviations from the linear theory predictions are less
than 5\%, then we smoothly transition $C_l(r)$ values to
$r_\up{lin}$ beyond which the adopted $C_l(r)$ values exactly become
the linear theory predictions. The redshift-space multipoles
$\xi_l(r)$ at $r\geq40\hmpc$ are then obtained by using 
equations~(\ref{eq:multipole}) with $C_l(r)$, in place of constant $C_l$,
the linear theory prediction. We simply set 
$r_\up{lin}=500\hmpc$, and $P_l(k)$ values are insensitive to the choice
of $r_\up{lin}$ as long as $r_\up{lin}>100\hmpc$.
We adopt the \citet{smith} prescription for the matter power spectrum 
$\pmm(k)$ and its Fourier transform
$\ximm(r)$ to compute $\xiR(r)$ on large scales,
and the asymptotic galaxy bias factor is computed by
\beeq
b_0={1\over {\bar n}_g}\int_0^\infty dM{dn\over dM}\langle N\rangle_M 
b_\up{h}(M),
\label{eq:bias}
\eneq
where $b_\up{h}(M)$ is the bias factor of halos of mass $M$. We use
the \citet{raby2} formulation with coefficients obtained by \citet{mtl},
which yield a better fit to the simulations.

\begin{figure}
\centerline{\epsfxsize=3.5truein\epsffile{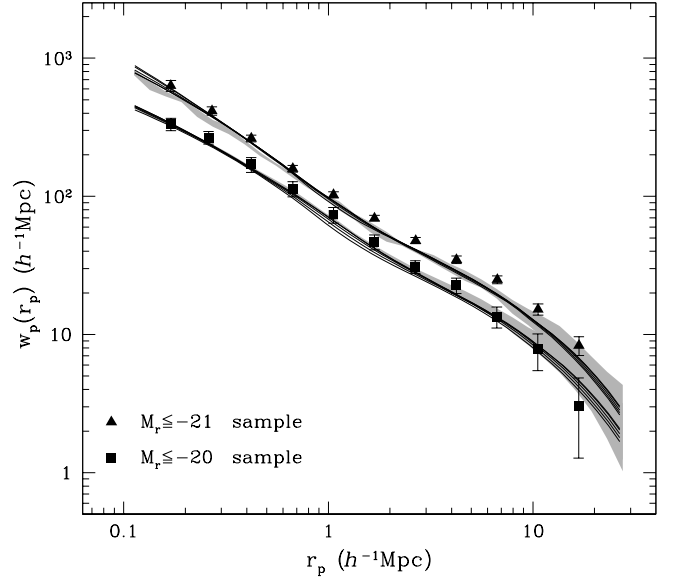}}
\caption{Projected correlation functions $\wpp$ for SDSS galaxy samples with
absolute-magnitude limits
$M_r\leq-21$ ({\it upper points/lines}) and $M_r\leq-20$ 
({\it lower points/lines}). Solid
lines represent the best-fit analytic model predictions of $\wpp$ 
for five different combinations of cosmological parameters,
with increasing $\rms$ and decreasing $\OM$ from lowest to highest curves
(see Table~\ref{tab:par}). Shaded regions show the statistical uncertainty of
the central cosmological model only, 
computed from the error on the mean of five independent $N$-body
simulations. For comparison, we plot the $\wpp$ measurements of the galaxy
samples from \citet{idit3}, with error bars that 
are the diagonal elements of the covariance matrix.
The model fits are acceptable when the full covariance matrix is
considered.}
\label{fig:wp}
\end{figure}

\section{Recovering the Real-Space Galaxy Power Spectrum}
\label{sec:gal}
Before turning to the bias between the galaxy power spectrum and the linear
matter power spectrum, we investigate how well the method 
used by \citet{max1,max2} 
recovers the true real-space galaxy power spectrum. Along the way, we test
the accuracy of our analytic model prediction for the redshift-space power 
spectrum against the results obtained from the $N$-body galaxy catalogs 
described in \S~\ref{sec:numerical}. Our basic approach to predicting the
galaxy power spectra is that we first determine HOD parameters for an 
observed galaxy sample given a cosmological model, then calculate the galaxy 
power spectra using this inferred relation between galaxies and dark
matter halos. We obtain HOD parameters
by fitting the mean number densities 
$\bar{n}_g$ and projected correlation function measurements $\wpp$ of the SDSS 
$M_r\leq-20$ and $M_r\leq-21$ 
galaxy samples, taking into account the full
covariance error matrix, estimated through jackknife resampling of the 
observational sample \citep{idit3}.

\begin{deluxetable*}{cccccccccc}
\tabletypesize{\scriptsize}
\tablewidth{0pt}
\tablecaption{HOD Parameters of the five $N$-body models}
\tablehead{\multicolumn{3}{c}{} & \multicolumn{3}{c}{$M_r\leq-20$} &
\colhead{} & \multicolumn{3}{c}{$M_r\leq-21$} \\ \cline{4-6} \cline{8-10}
\colhead{Model} & \colhead{$\Omega_m$} & \colhead{$\sigma_8$}
& \colhead{$M_\up{min}(\hmsun)$} & \colhead{$M_1(\hmsun)$}
& \colhead{$M_\up{cut}(\hmsun)$} & & \colhead{$M_\up{min}(\hmsun)$} & 
\colhead{$M_1(\hmsun)$} & \colhead{$M_\up{cut}(\hmsun)$}}
\startdata
1 & 0.10 & 0.95 & $2.95\times10^{11}$ & $5.37\times10^{12}$ & 
$1.40\times10^{13}$ & & $1.64\times10^{12}$ & $2.05\times10^{13}$ & 
$2.89\times10^{13}$ \\
2 & 0.16 & 0.90 & $4.80\times10^{11}$ & $8.03\times10^{12}$ & 
$1.37\times10^{13}$ & & $2.60\times10^{12}$ & $3.06\times10^{13}$ & 
$3.47\times10^{13}$ \\
3 & 0.30 & 0.80 & $9.33\times10^{11}$ & $1.32\times10^{13}$ & 
$1.38\times10^{13}$ & & $4.83\times10^{12}$ & $4.82\times10^{13}$ & 
$4.90\times10^{13}$ \\
4 & 0.47 & 0.69 & $1.44\times10^{12}$ & $1.62\times10^{13}$ & 
$1.78\times10^{13}$ & & $7.15\times10^{12}$ & $5.69\times10^{13}$ & 
$6.53\times10^{13}$ \\
5 & 0.63 & 0.60 & $1.89\times10^{12}$ & $1.60\times10^{13}$ &
$2.46\times10^{13}$ & & $8.91\times10^{12}$ & $5.65\times10^{13}$ & 
$6.85\times10^{13}$ \\
\enddata
\tablecomments{The HOD parameters of the five $N$-body models are determined to
reproduce the same clustering $\wpp$
of the SDSS galaxy samples with $M_r\leq-20$ 
and $M_r\leq-21$, and to match the number densities 
$\bar n_g=5.74\times10^{-3}~(\hmpc)^{-3}$ for the $M_r\leq-20$ sample and
$\bar n_g=1.17\times10^{-3}~(\hmpc)^{-3}$ for the $M_r\leq-21$ sample, 
respectively.}
\label{tab:par}
\end{deluxetable*}

Figure~\ref{fig:wp} shows the projected correlation functions of the two
SDSS galaxy samples, where the solid lines are the analytic model predictions
for the five different cosmological models (listed in Table~\ref{tab:par}),
obtained by fitting the \citet{idit3} measurements shown as symbols.
The error bars show only the diagonal elements of the covariance matrix of
the $\wpp$ measurements, and the analytic model fits to the measurements are
acceptable over a wide range of parameter combinations ($\OM$, $\rms$)
when the full covariance matrix is considered, since the errors 
between data points are strongly correlated. 
However, there is strong degeneracy between the shape $\Gamma$ and spectral
index $n_s$ of the power spectrum, and the best-fit HOD parameters are 
insensitive to $\Gamma$ and $n_s$, as discussed in \S~\ref{sec:uncer} below.
Changing the adopted value of $z_\up{max}$ has little effect on the HOD
parameters inferred by fitting $\wpp$, though it affects the $\chi^2$ values
of these fits slightly. We use $z_\up{max}=40\hmpc$ for further analyses.
The shaded regions represent the statistical uncertainty in the mean value of 
$\wpp$, computed from the dispersion among the five independent populated
$N$-body simulations, for the central model
($\OM=0.3$, $\rms=0.8$; see Table~\ref{tab:par}).
Our analytic model predictions for $\wpp$ agree with
the $N$-body results within 5\% fractional differences. 

\begin{figure}[t]
\centerline{\epsfxsize=3.5truein\epsffile{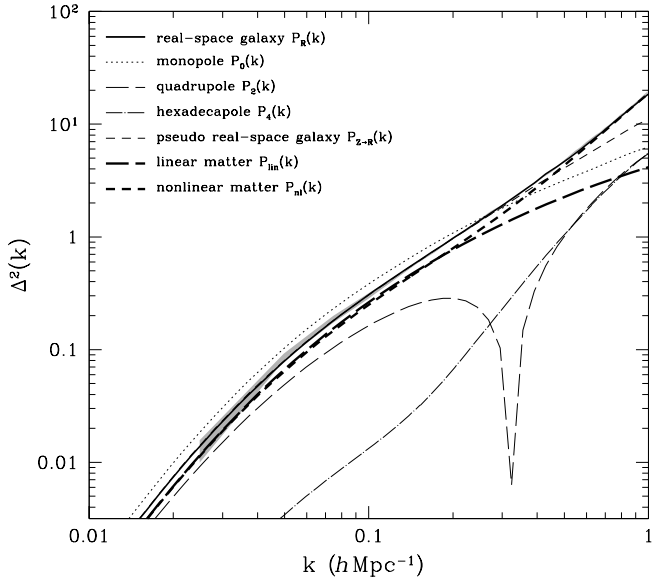}}
\caption{Dimensionless real-space and redshift-space multipole power spectra
$\dimp$ for the HOD parameters appropriate to
$M_r\leq-20$ galaxies. Various curves represent
the analytic predictions of the central model
for the corresponding galaxy power spectra indicated in
the legend, and shaded regions show the statistical uncertainty
on the real-space galaxy power spectrum computed from the five $N$-body 
simulations. The light, short dashed curve, labeled $\pzr(k)$ (see, 
eq.[\ref{eq:recov}]), is a linear combination of redshift-space multipoles
that reduces to $\pR(k)$ in linear regime; it is largely obscured by the solid
curve. Thick dashed curves
represent the linear ($long$) and nonlinear ($short$) matter power spectra. 
Note that the quadrupole $P_2(k)$ crosses zero at $k=0.3\hmpci$, and
$-P_2(k)$ is plotted at larger $k$.}
\label{fig:multi}
\end{figure}

\begin{figure}[t]
\centerline{\epsfxsize=3.4truein\epsffile{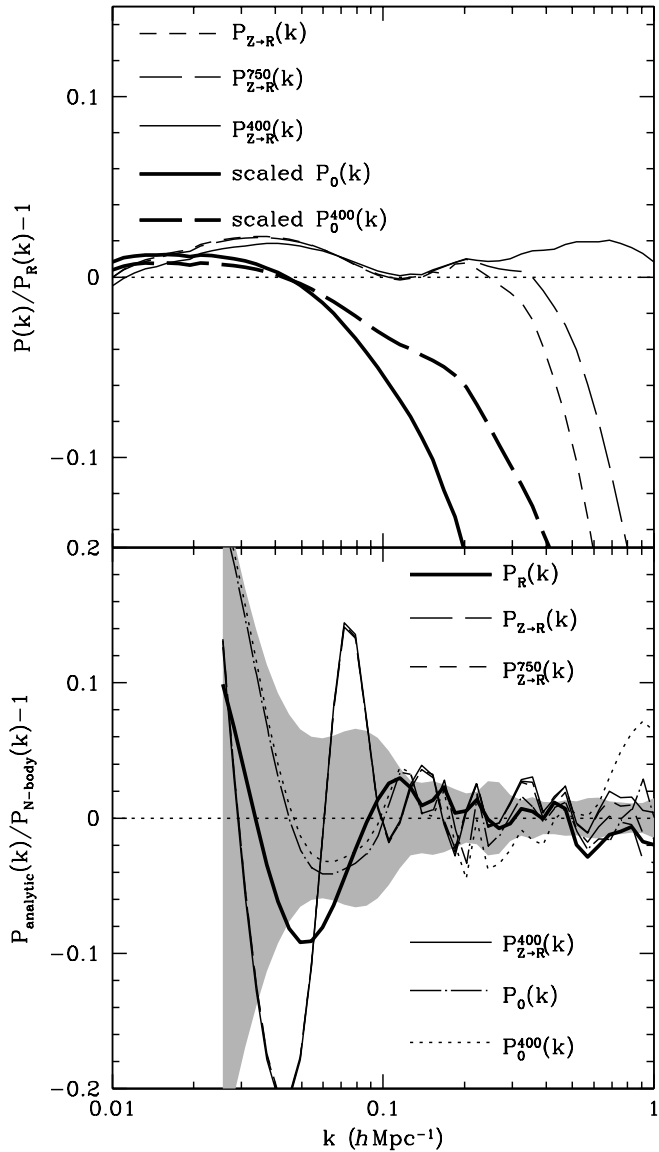}}
\caption{The upper panel plots pseudo real-space galaxy power spectra and 
redshift-space monopoles with and without Finger-of-God (FoG)
compression relative to the analytic model prediction for $\pR(k)$,
assuming HOD parameters that fit the SDSS galaxy sample with $M_r\leq-20$.
Thresholds for the 
FoG compression are 400~$\kms$ and 750~$\kms$, as indicated in the
legend. The redshift-space monopoles are scaled by a constant factor 
predicted by linear theory to match $\pR(k)$ at large scales.
The bottom panel shows the fractional difference
between the analytic model calculations and simulation results for the
corresponding galaxy
power spectra. Shaded regions represent fractional statistical
uncertainty on $\pR(k)$ from the $N$-body simulations caused by the finite
simulation volume. The other power spectra
have larger statistical uncertainties because of finite volume effects on
redshift-space distortions. Note that $\pzr(k)$ with and without FoG
compression follow each other too closely on large scales to be separated
from the thin solid line.}
\label{fig:comp}
\end{figure}

\begin{figure*}[t]
\centerline{\epsfxsize=6.0truein\epsffile{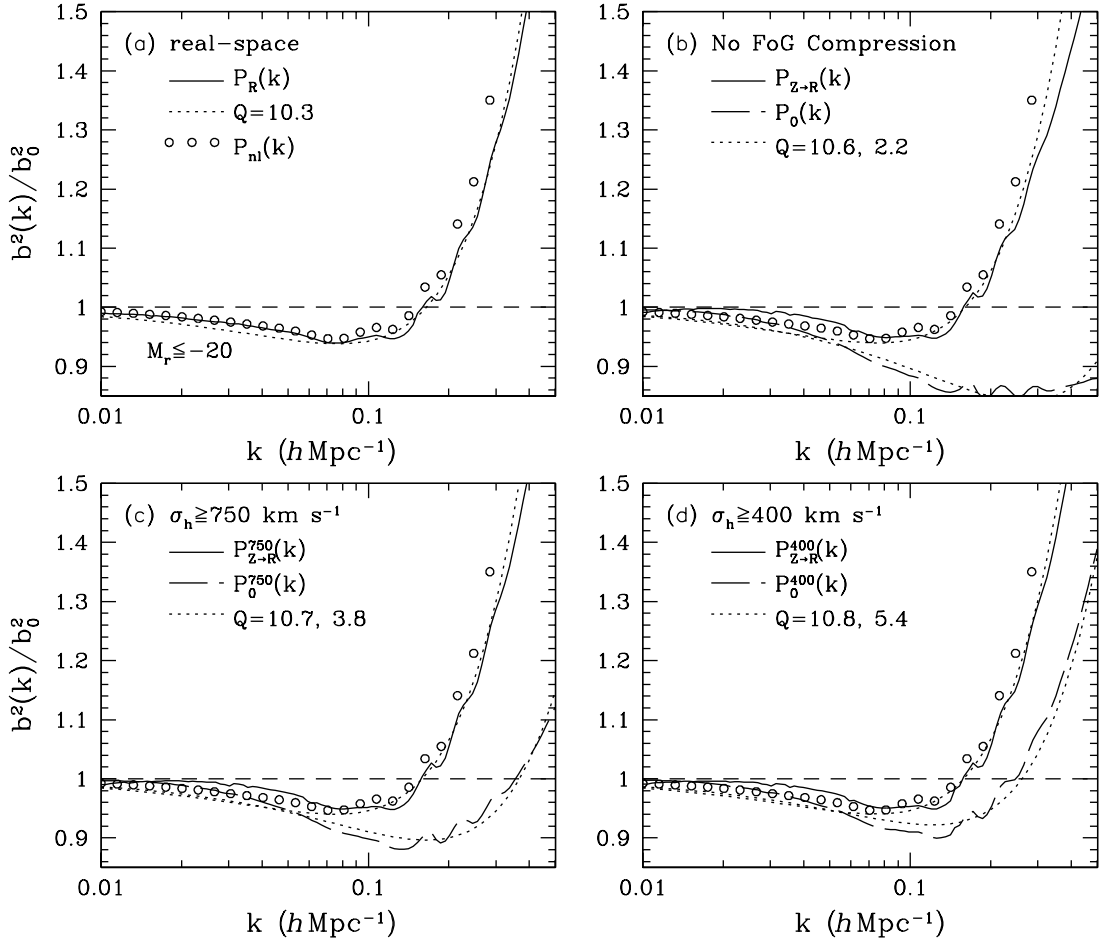}}
\caption{Scale-dependent bias functions calculated for the $M_r\leq-20$ galaxy
sample. Panel~($a$) shows the real space $\pR(k)$, and panels~($b$) to ($d$)
show $\pzr(k)$ and $P_0(k)$ with varying levels of FoG compression.
In each panel, circles show the ratio of the nonlinear matter power spectrum
to the linear matter power spectrum, and dotted lines show the $Q$-model
curves that best reproduce our predicted $b^2(k)$.}
\label{fig:shape}
\end{figure*}

\begin{figure*}[t]
\centerline{\epsfxsize=6.0truein\epsffile{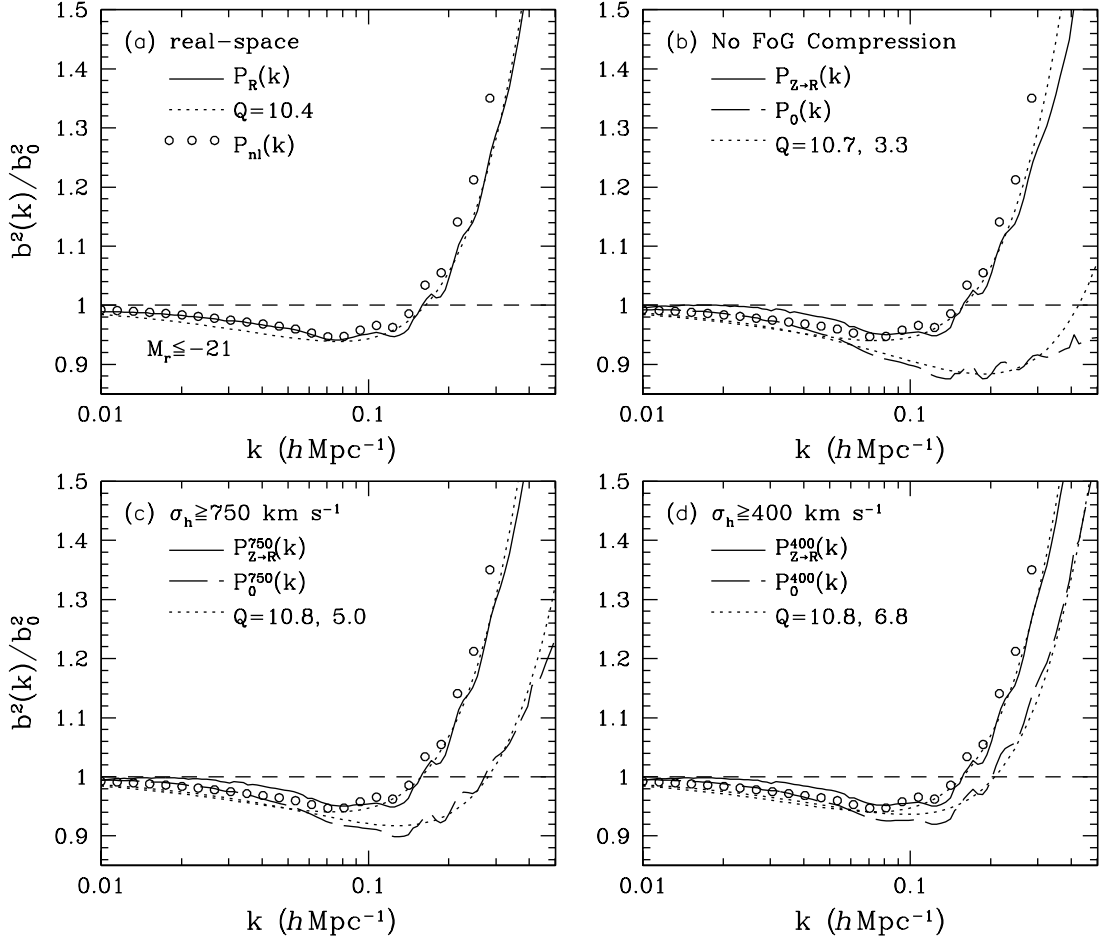}}
\caption{Same as Fig.~\ref{fig:shape}, but for the $M_r\leq-21$ sample.}
\label{fig:shape21}
\end{figure*}

\begin{figure*}[t]
\centerline{\epsfxsize=6.0truein\epsffile{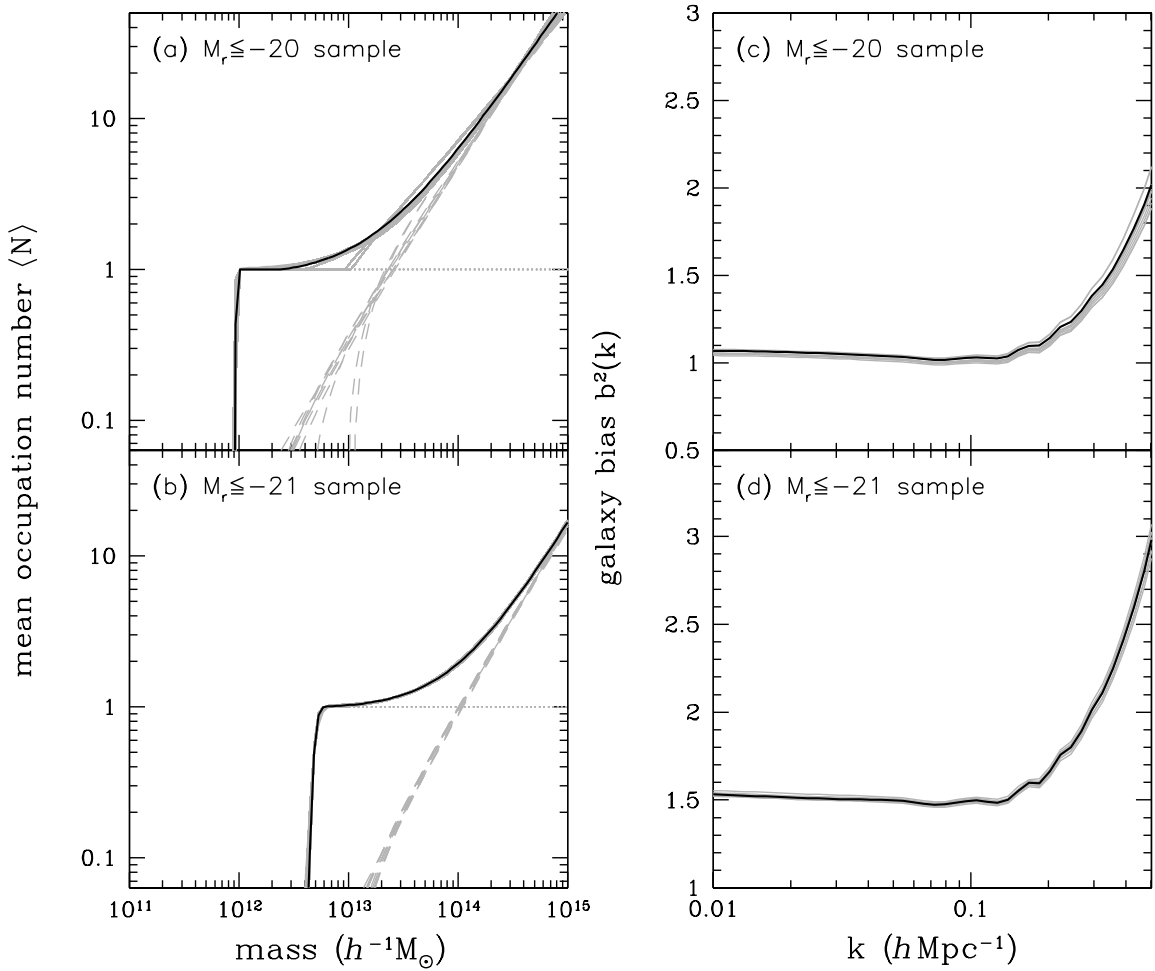}}
\caption{Impact of HOD parameter uncertainties on bias shape.
We compare our best-fit HOD models ($solid$) to ten HOD models ($gray$) for
each galaxy sample that have $\Delta\chi^2\leq1$ relative to the best-fit
models, randomly chosen from the Monte Carlo Markov Chain.
Left-hand panels show mean occupation functions and right-hand panels
show $b^2(k)$.}
\label{fig:unc}
\end{figure*}

\begin{figure*}[t]
\centerline{\epsfxsize=6truein\epsffile{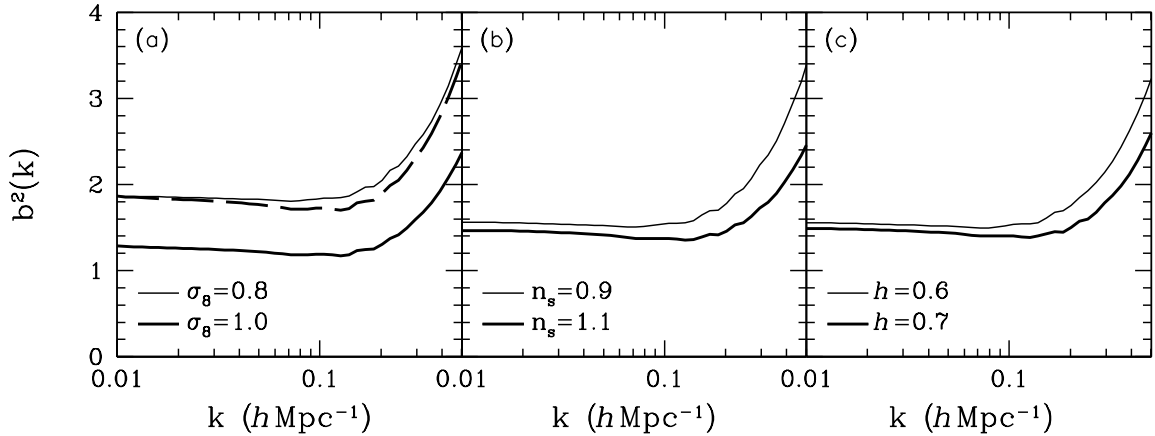}}
\caption{$b^2(k)$ for cosmological model variations. For each model in the
legend, we adjust HOD parameters to obtain the minimum $\chi^2$-fit to $\wpp$
observations. We then compute $\pR(k)$ and $b^2(k)=\pR(k)/\plin(k)$. The
dashed curve in the left panel shows $b^2(k)$ for the model with $\rms=1.0$
scaled to match the large-scale bias factor of the $\rms=0.8$ model.}
\label{fig:var}
\end{figure*}

\begin{figure}[t]
\centerline{\epsfxsize=3.7truein\epsffile{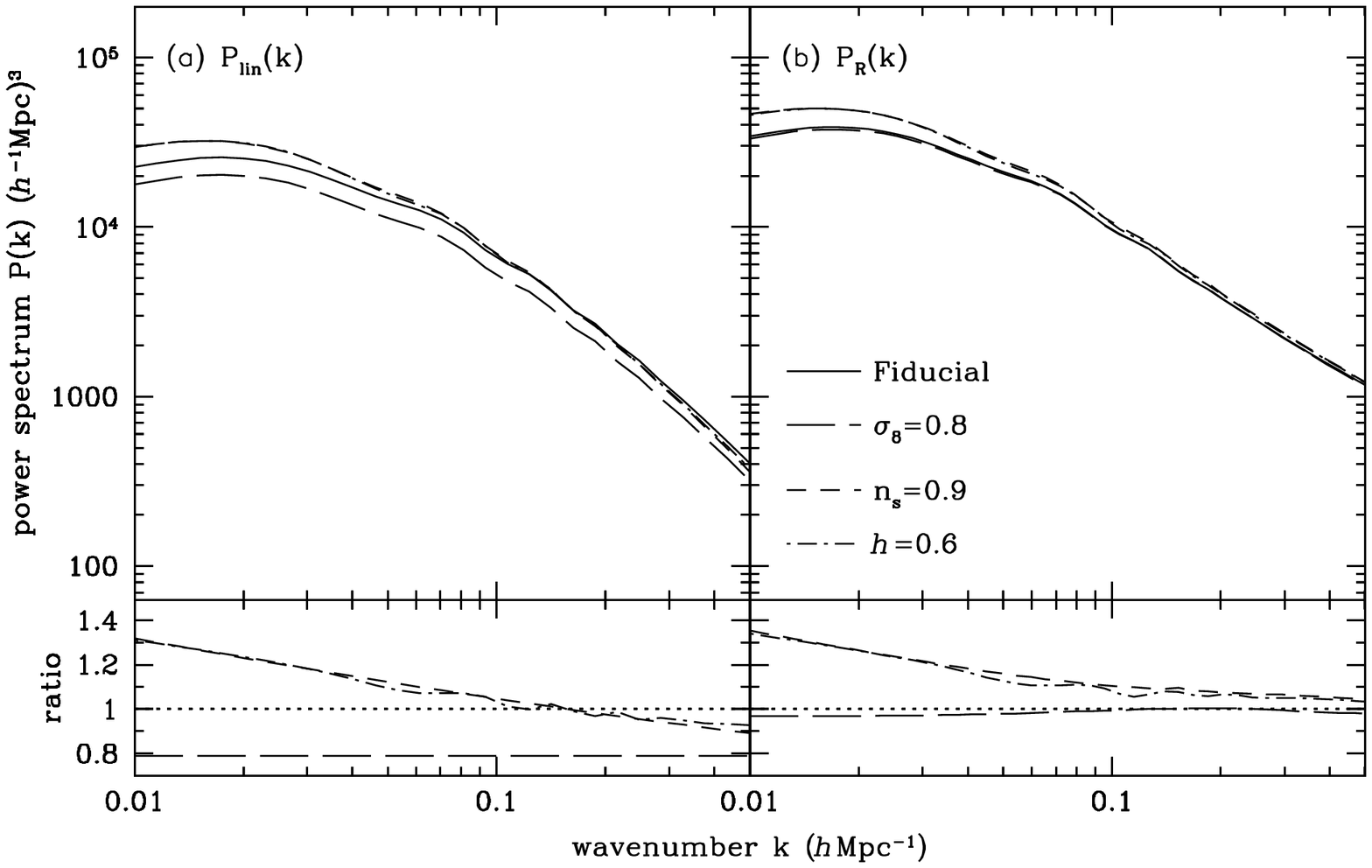}}
\caption{Effect of cosmological parameter variations on 
the linear matter ({\it left panel})
and observable galaxy ({\it right panel}) power spectra. The bottom panels
show the variations relative to the fiducial model power spectrum
($\rms=0.9$, $n_s=1.0$, $h=0.7$).}
\label{fig:dev}
\end{figure}

Using the analytic model and assuming the central cosmological
model, we illustrate the
dimensionless power spectra $\dimp=k^3P(k)/2\pi^2$ of the $M_r\leq-20$ galaxy 
sample in Figure~\ref{fig:multi}, where the thick solid line represents the
real-space galaxy power spectrum $\pR(k)$ and the shaded region is the 
statistical uncertainty in the mean value of $\pR(k)$. The finite box size
of the simulations puts a limit on Fourier modes
$k\geq k_\up{box}\equiv2\pi/L_\up{box}=0.025\hmpci$ that we can measure from
the populated halo catalogs. We test our analytic model predictions for galaxy
power spectra below. The two thick dashed lines are the matter power spectra
of the assumed cosmological model; the long-dashed line is the linear matter
power spectrum $\plin(k)$ and the short-dashed line is the nonlinear matter
power spectrum $\pnl(k)$, computed by using the \citet{smith} prescription.
On large scales, $\pR(k)$ has the some shape as $\plin(k)$ with 
normalization differing by $b_0^2$, the square of the large-scale
galaxy bias factor, as predicted by
the linear bias approximation (eq.[\ref{eq:bias0}]). However, on scales
$k\gtrsim0.2\hmpci$, where $\pnl(k)$ departs from $\plin(k)$ or 
$\Delta^2_\up{lin}(k)\simeq1$, $\pR(k)$ follows more closely the $\pnl(k)$
shape than the $\plin(k)$ shape.

We also present the analytic model predictions for the redshift-space 
multipole power spectra $\Delta^2_l(k)=k^3P_l(k)/2\pi^2$.
The dotted, long-dashed, and dot-dashed lines in Figure~\ref{fig:multi}
represent the redshift-space
monopole $P_0(k)$, quadrupole $P_2(k)$, and hexadecapole $P_4(k)$, 
respectively. On large scales, coherent peculiar velocities 
produce redshift-space distortion,
where the overdense regions
shrink and the underdense regions inflate in redshift-space. Therefore, when
averaged over angle, $P_0(k)$ is larger than $\pR(k)$. However, on small 
scales, nonlinear collapse and
random motions in virialized objects stretch systems along the 
line-of-sight, giving rise to the Finger-of-God (FoG) effect, which inflates
overdense regions and depresses their density contrast. Therefore, 
$P_0(k)$ is smaller than $\pR(k)$ on small scales.

In the linear regime, the redshift-space power spectrum can be written as
$P(k,\mu)=(1+\beta\mu^2)^2\pR(k)$, and the real-space power spectrum can
be reconstructed by using a linear combination of the three redshift-space 
multipoles to remove the unknown variable $\beta$,
\beeq
\pzr(k)\equiv P_0(k)-{1\over2}P_2(k)+{3\over8}P_4(k),
\label{eq:recov}
\eneq
which exactly reduces to $\pR(k)$ if the linear theory approximation holds
\citep{nick2}. 
The upper panel of Figure~\ref{fig:comp} tests the ability of 
equation~(\ref{eq:recov}) to recover the true $\pR(k)$.
$\pzr(k)$ recovers $\pR(k)$ to 5\% at $k\leq0.4\hmpci$, while it 
substantially underestimates $\pR(k)$ at $k>0.4\hmpci$.
This pseudo real-space power spectrum $\pzr(k)$
can be used to correct for
the effect of redshift-space distortions and to estimate $\pR(k)$. 
A $k$-by-$k$ application of equation~(\ref{eq:recov}) is, essentially, the
procedure referred to as 
{\it the disentanglement approach} in \citet{max1,max2},
and our $\pzr(k)$ corresponds to $P_{gg}(k)$ in their notation. Their
alternative, {\it modeling approach}, is to construct two more power spectra,
namely $P_{gv}(k)$ and $P_{vv}(k)$, from redshift-space multipole measurements,
then solve for $\pR(k)$ with the linear theory approximation 
\citep{max1,max2}. The former method gives a more robust approximation to
$\pR(k)$ because of the cancellation
of deviations in the multipoles from the linear theory approximation on
nonlinear scales, but it yields larger statistical errors because it
effectively marginalizes over uncertainties in the other two power spectra
at each $k$. In contrast, the modeling approach
provides decorrelated estimates of $\pR(k)$ with smaller error
bars, but it is subject
to systematic error in the linear theory approximation when modeling the other
two power spectra. \citet{2dfn} use the Fourier-based method of
\citet{pvp}, which extends the \citet{feldman} method to include 
luminosity-dependent 
galaxy bias, computing the monopole $P_0(k)$ in our notation. We also
take $P_0(k)$ as one of our principal estimates of $\pR(k)$.

A significant part of the
nonlinearity in redshift-space distortions arises from 
the random motion of galaxies in virialized objects, which can be identified by
finding groups or clusters. The FoG effects can be removed by moving all 
galaxies in the same virialized object to the same redshift.
In practice, clusters of galaxies are identified by applying the
friends-of-friends algorithm to galaxy positions in redshift-space. Two 
galaxies are assumed to belong to the same cluster if the density windowed
through an ellipse (usually several 
times longer in the radial than transverse direction)
is higher than an overdensity threshold, and the radial dispersion of member
galaxy positions is set equal to the transverse dispersion to compress the
FoG effect, if the former exceeds the latter \citep{max1,max2}. Here we
assume a perfect process of compressing the FoG effects; all the halos with
velocity dispersion $\sigh$ greater than a threshold are compressed by setting
$\sigh=100~\kms$, and none of the halos with $\sigh$ smaller than the threshold
are affected. The pseudo real-space power spectrum $\pzrF{750}(k)$ with FoG 
compression threshold $\sigh=750~\kms$ agrees with $\pR(k)$ to 5\%
at $k\leq0.5\hmpci$, and these halos are easily identifiable in practice.
With a more aggressive threshold $\sigh=400~\kms$, $\pzrF{400}(k)$ recovers
$\pR(k)$ to 2\% at $k\leq1\hmpci$. We also compare the $P_0(k)$ shape to 
$\pR(k)$ after scaling the constant factor $C_0$ in \S~\ref{sec:redmult}.
The scaled $P_0(k)$ is depressed by 5\% at $k=0.1\hmpci$ compared to $\pR(k)$.
FoG compression helps at $k\simeq0.1\hmpci$, 
but the difference between $P_0^{400}(k)$ and $\pR(k)$ reaches 5\% at 
$k\simeq0.2\hmpci$ and 10\% at $k\simeq0.3\hmpci$.

Having shown the agreement between the analytic model predictions and the
$N$-body results in $\wpp$, we now test the accuracy of the analytic model
predictions for real-space and redshift-space power spectra against the 
$N$-body simulations.
The bottom panel shows the fractional difference in galaxy power spectra
between the analytic model and the $N$-body results, where the shaded region 
shows only the statistical uncertainty in the mean value of $\pR(k)$ from
the simulations. Note that the uncertainties on the other power spectra are 
larger and are not shown. The analytic model calculation of $\pR(k)$ is 
accurate to
better than a few percent at $k>0.08\hmpci$, while it is difficult to assess
the statistical significance at $k<0.08\hmpci$, where the simulations only
have few Fourier modes due to the finite box size. 
Since linear theory should become accurate on large scales, it would be 
surprising if the analytic model became less accurate on this regime.
Our analytic model also
provides accurate predictions for $P_0(k)$ and $\pzr(k)$, both with and without
FoG compression, at $k>0.1\hmpci$.

\section{Recovering the Linear Matter Power Spectrum}
\label{sec:lin}
We now turn to our principal results, the scale-dependent bias relation
$b^2(k)=\pobs(k)/\plin(k)$ between observable galaxy power spectra 
$\pobs(k)$ and the
linear matter power spectrum $\plin(k)$. As potentially observable power 
spectra, we consider $\pR(k)$ (inferred from the angular clustering power 
spectrum), $\pzr(k)$,
and $P_0(k)$, with varying levels of FoG compression for the latter two.
Here we use HOD constraints for the $M_r\leq-20$ and $M_r\leq-21$ samples
of \citet{idit3} based on their $\wpp$ measurements from SDSS Data Release~2.
These could be further
improved with $\wpp$ measurements from subsequent SDSS data and
with constraints from the group multiplicity function \citep{multip}.
We investigate the uncertainties in the power spectrum recovery 
associated with our HOD modeling and with variations of the assumed 
cosmological model. We use a
{\scriptsize{CMBFAST}} transfer function \citep{cmbfast} 
to compute $\plin(k)$ for a given cosmology, in place of the \citet{george}
parametrization used in \S~\ref{sec:gal}
to test our analytic model against the $N$-body 
simulations (which used these initial conditions). Unless explicitly stated
otherwise, we use a cosmological model with $\OM=0.3$, $\OD=0.7$, $n_s=1.0$,
$h=0.7$, $\OB h^2=0.02$, and $\rms=0.9$ in this section.

\subsection{Scale-Dependent Bias of the $M_r\leq-20$ and $M_r\leq-21$
Galaxy Samples}
Figure~\ref{fig:shape} plots the scale-dependent bias functions 
$b^2(k)/b_0^2$ of the galaxy
sample with $M_r\leq-20$, in which $b_0$ is the asymptotic
galaxy bias factor computed via equation~(\ref{eq:bias}).
Figure~\ref{fig:shape}$a$ shows the real-space $\pR(k)$ ({\it solid line}) 
and the nonlinear matter power spectrum $\pnl(k)$ ($circles$), which follow each
other remarkably closely. A suppression of $\pnl(k)$ at $k\simeq0.1\hmpci$
relative to $\plin(k)$ results from the nonlinear damping of linear 
perturbations. The physical 
origin of the $\pnl(k)$ shape is discussed by \citet{smith2006}. 
Note that we compute the nonlinear matter power spectrum $\pnl(k)$ using the
\citet{smith} prescription, modified to utilize a {\scriptsize{CMBFAST}} 
transfer function. $\plin(k)$ retains the baryonic acoustic oscillations (BAO)
imprinted by sound waves in
the baryon-photon plasma before recombination, but nonlinear evolution washes
out the oscillations at higher $k$ (see, \citealt{danielbao}).
The bias functions in Figure~\ref{fig:shape} exhibit small oscillations at 
$k>0.1\hmpci$ because the smoother nonlinear power spectra are divided by
a reference $\plin(k)$ that retains the BAO features at their original 
strength.

When normalized to the large-scale amplitude, the real-space power 
spectrum of $M_r\leq-20$ galaxies falls below $\plin(k)$ by $\simeq5\%$ at
$k\simeq0.1\hmpci$, climbs above by 5\%
at $k\simeq0.22\hmpci$, and rises rapidly thereafter. However, for this galaxy
sample, the assumption that $\pR(k)$ traces the
$nonlinear$ matter power spectrum remains quite accurate, to 1\% 
at $k\leq0.2\hmpci$ and 7\% at $k=0.3\hmpci$.
The dotted curve shows the $Q$-model prediction of equation~(\ref{eq:qmodel}) 
with $Q=10.6~(\hmpc)^2$, 
which gives a least-squares fit to our predicted bias curve 
over the range
$0.01\hmpci<k<0.3\hmpci$.  The largest difference is at $k\simeq0.04\hmpci$,
where the $Q$-model predicts a 5\% deviation from $\plin(k)$ while we find 3\%.
The $Q$-model accurately traces the predicted bias shape beyond $k=0.1\hmpci$.
However, if one normalized the curves to match at $k\simeq0.04\hmpci$,
which might well happen in practice because of the large
statistical uncertainties at low $k$, then deviations
in the bias shape would be smaller at $k<0.1\hmpci$ and larger
at $k>0.1\hmpci$.

Figure~\ref{fig:shape}$b$ plots $\pzr(k)$ and $P_0(k)$
with no FoG compression. The bias shape for $\pzr(k)$ is
qualitatively similar to
the true real-space $\pR(k)$, but it follows the $\plin(k)$ shape more 
closely than the $\pnl(k)$ shape at $k\leq0.05\hmpci$.
$\pzr(k)$ is poorly described by the $Q$-model prescription, with differences 
of 6\% at $k\simeq0.05\hmpci$. The bias shape for $P_0(k)$ is completely 
different from those for $\pR(k)$ and $\pzr(k)$. By $k=0.1\hmpci$, 
$P_0(k)/b_0^2$ is below $\plin(k)$ by 10\% and $\pnl(k)$ by 5\%, and it
remains below $\plin(k)$ by 10-15\% to $k=0.5\hmpci$. The best-fit $Q$-model
has $Q=2.2~(\hmpc)^2$, and it has several percent differences from our $P_0(k)$
prediction at $k\leq0.1\hmpci$.

Figures~\ref{fig:shape}$c$ and \ref{fig:shape}$d$ illustrate the effects of 
FoG compression on $\pzr(k)$ and $P_0(k)$. Changes to $\pzr(k)$ are minor,
though the agreement with the true $\pR(k)$ is significantly improved
at $k\geq0.25\hmpci$. Changes to the monopole $P_0(k)$ are much more
substantial, especially at $k\geq0.2\hmpci$; suppression of $P_0(k)$ by
the virial motions of satellites is much stronger than for $\pzr(k)$.
However, the shape of $P_0(k)$ remains quite far from that of 
the true $\pR(k)$, 
even for the aggressive FoG compression with $\sigh\geq400~\kms$.

Figure~\ref{fig:shape21} plots the scale-dependent bias functions of the galaxy
sample with $M_r\leq-21$, in the same format as Figure~\ref{fig:shape}.
Compared to the $M_r\leq-20$ galaxy sample, this galaxy sample 
has a large-scale bias factor higher by 20\%, and hence a power spectrum
amplitude higher by nearly 50\%. However, the shape of $b^2(k)/b^2_0$
is nearly identical for the two galaxy samples
over a large dynamic range $k\leq0.3\hmpci$, as one can see by using the
nonlinear matter power spectrum ($circles$) as a reference.
In each case, the best-fit $Q$-model reproduces our predictions for 
$\pR(k)$, $\pzr(k)$, and $P_0(k)$ at roughly the 5\% level for 
$k\leq0.3\hmpci$. 

\subsection{Impact of HOD Uncertainties}
\label{sec:uncer}
Our method for calculating $\pobs(k)$ and then 
$b^2(k)$ is that we first determine HOD parameters
by fitting $\wpp$ measurements given a cosmological model.
With perfect knowledge of HOD and cosmological parameters,
it should be possible in principle to
calculate $\pobs(k)$ exactly and fit the
observed galaxy power spectrum to constrain $\plin(k)$ up to high $k$.
However, uncertainties in the parameters result in uncertainty in
our $\pobs(k)$ calculations. Therefore, the dependence of $\pobs(k)$ on
{\it our adopted fiducial HOD model} 
will increase uncertainty in cosmological
parameter estimation relative to measurement errors alone.

To investigate these uncertainties,
we adopt the 5-parameter HOD model of \citet{zheng3} 
instead of the 3-parameter model described in \S~\ref{sec:numerical},
so that we do not underestimate HOD uncertainties because of an overly
restrictive form. We determine best-fit parameters using this parametrization
and the {\scriptsize{CMBFAST}} power spectrum. Then we generate
a Markov Chain Monte Carlo based on the Metropolis-Hastings algorithm
by fitting the $\wpp$ measurements with covariance matrix. In this 
parametrization, the mean 
occupation function for central galaxies is
\beeq
\langle N_\up{cen}\rangle_M={1\over2}\left[1+\up{erf}\left({\log M-\log\Mmin
\over \sigma_\up{\log M}}\right)\right],
\eneq
where erf$(x)$ 
is the error function, which corresponds to a Gaussian scatter of width
$\sigma_\up{\log M}$ in $\log L$ at fixed halo mass \citep{zheng6}.
The sharp cut-off
of \S~\ref{sec:numerical} is the limiting case of 
$\sigma_\up{\log M}\rightarrow0$. The distribution of $N_\up{cen}$
about the mean is a nearest-integer or Bernoulli distribution.
The mean occupation function for
satellite galaxies is
\beeq
\langle N_\up{sat}\rangle_M=\left({M-M_0\over M'_1}\right)^\alpha,
\eneq
for $M>M_0$, and halos of $M\leq M_0$ are devoid of satellite
galaxies, so the slope of the satellite occupation is now free instead of
fixed to one as in \S~\ref{sec:numerical}. The number of satellites is
Poisson-distributed about $\langle N_\up{sat}\rangle_M$. 

In Figure~\ref{fig:unc}, the left-hand panels show the mean occupation functions
of ten HOD models for each sample that have $\Delta\chi^2\leq1$ relative to
the best-fit models, randomly selected from the Markov Chain.
The HODs are tightly constrained by the $\wpp$ data, especially for the
$M_r\leq-21$ sample. The right-hand panels show $b^2(k)$ of the ten HOD
models. Here we present the HOD dependence of $\pobs(k)$ in terms of $b^2(k)$
since $\plin(k)$ is identical for all the models, and we only show 
$b^2(k)$ for $\pR(k)$, noting that FoG effects are similar
for the fixed cosmological parameters. Changing HOD parameters in the 
1-$\sigma$ allowed range alters $b^2(k)$ by up to 4\% at $k=0.5\hmpci$ for
the $M_r\leq-20$ sample, and by $\leq2\%$ for $k\leq0.2\hmpci$. Induced
errors for the $M_r\leq-21$ sample are even smaller, by about a factor of
two, because of the smaller measurement errors in $\wpp$. 
Hence its impact on 
cosmological parameter constraints is small at $k\leq0.2\hmpci$, and we 
discuss the impact as we include measurements at higher $k$ in \S~\ref{sec:lrg}.
Note that even
these small uncertainties in $\pobs(k)$ associated with HOD parameter
uncertainties can be reduced using $\wpp$ measurements from larger SDSS 
samples, and by bringing in additional constraints such as the group 
multiplicity function. These uncertainties are
unlikely to make a major contribution to overall error 
budget in $\plin(k)$ recovery.

Our model fitting assumes that the galaxy HOD is independent of large-scale
environment. However, recent studies show that the clustering of galaxy mass
halos has a substantial dependence on halo formation time as well as halo
mass (e.g., \citealt{gao,harker}), opening the door to environment-dependent
HODs. \citet{croton} use the Millennium simulation of \citet{mille}
to investigate the effect of environment-dependent HODs on galaxy clustering
in semi-analytic
galaxy models and find that environmental effects change galaxy bias factors
for luminosity thresholded samples by a few percent, but with virtually 
no scale-dependence (see also \citealt{zhu}).
Our HOD parameters are insensitive to the large-scale
amplitude of $\wpp$, being
driven mainly by shape of $\wpp$ in a regime where the
contributions of galaxy pairs
from single halos and two distinct halos are comparable.
Therefore, we suspect that environment dependence
at the level predicted by \citet{croton} would have few percent impact
on the predicted asymptotic galaxy bias factor
$b_0^2$ but probably much smaller effect on the scale-dependence
of $b^2(k)/b_0^2$. Furthermore, studies of void probabilities \citep{void}
and color-separated correlation functions in ``scrambled'' group catalogs
\citep{blanton} suggest that the \citet{croton} models overpredict the
environmental dependence of galaxy HODs.
Nevertheless, the environment-dependent issue merits
further investigation in future work, as the statistical precision of the
measurements themselves sets better.

\subsection{Sensitivity of Galaxy Power Spectrum to $\plin(k)$}
In approach to real observations, we will need to simultaneously fit 
cosmological parameters and HOD parameters using $\pobs(k)$ and $\wpp$ as
constraints and to marginalize over HOD parameters. The ability to constrain
cosmological parameters depends on the measurement errors in $\pobs(k)$ and on
the sensitivity of $\pobs(k)$ to $\plin(k)$. Here we investigate the sensitivity
of $\pobs(k)$ to $\plin(k)$ given the constraints of $\wpp$. We discuss the
impact on cosmological parameter estimation in \S~\ref{sec:lrg}.

In Figures~\ref{fig:var} and~\ref{fig:dev}, we illustrate the changes in 
$\pobs(k)$ and $b^2(k)$ for cosmological model variations using the 
$M_r\leq-21$ sample. We adjust HOD parameters to fit $\wpp$ with
each cosmological model shown in the legend. Figure~\ref{fig:var}$a$ plots
$b^2(k)$ for models with $\rms=0.8$ and $\rms=1.0$. Note that for specified
cosmology, fitting $\wpp$ completely determines both the shape and
{\it amplitude} of $b^2(k)$. The thick dashed curve
shows $b^2(k)$ for the $\rms=1.0$ model scaled to have the same large-scale
bias factor as the $\rms=0.8$ model.
The few percent difference in scaled $b^2(k)$ at $k\simeq0.1-0.2\hmpci$
results from the shape variation in the \citet{smith}
$\pnl(k)$ prescription for different $\rms$ values.

\begin{figure}[t]
\centerline{\epsfxsize=3.5truein\epsffile{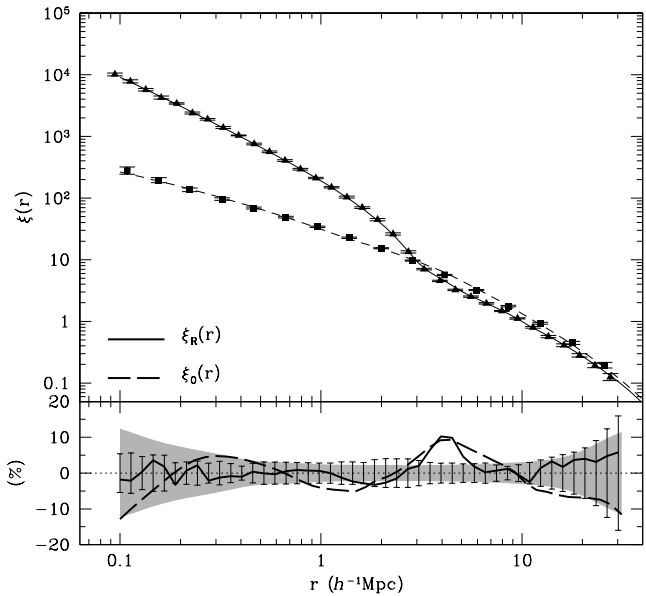}}
\caption{Test of analytic model predictions for real-space $\xiR(r)$ and 
redshift-space monopole $\xi_0(r)$. Here we take the HOD parameters for the
luminous red galaxy (LRG) sample from \citet{zheng5} as input and 
populate one large volume $N$-body simulation at $z=0$. Symbols 
represent measurements of $\xiR(r)$ and $\xi_0(r)$ from the
simulation and curves represent analytic model predictions. The
attached bottom panel shows the fractional differences between the analytic
model calculations and the simulation results. Statistical uncertainties are
computed by jackknife resampling of the eight octants of the simulation box
and shown as error bars for $\xiR(r)$ and shaded regions for $\xi_0(r)$.}
\label{fig:lrgtest1}
\end{figure}

Figures~\ref{fig:var}$b$ and~\ref{fig:var}$c$ plot two more model sequences
with variations in $n_s$ and~$h$, which change the shape of $\plin(k)$ in 
similar but not identical ways. Roughly speaking, fitting
$\wpp$ fixes the $\pobs(k)$ amplitude at $k\simeq0.5\hmpci$ for the
$M_r\leq-21$ sample. The large-scale bias factors are therefore slightly
different for distinct $n_s$ or~$h$ values. At $k\leq0.1\hmpci$, the
shapes of $b^2(k)$ are nearly identical and nearly scale-independent, among
the $n_s$ and~$h$ model sequences. On smaller scales, the $b^2(k)$ curves
separate as HOD parameters adjust to try to produce similar $\wpp$ from
models with different matter power spectra.

Figure~\ref{fig:dev} plots $\plin(k)$ and $\pR(k)$ in the upper panels and
their ratios relative to the fiducial model power spectra in the bottom panels,
for the same model sequences considered in Figure~\ref{fig:var}. 
Thus, this figure shows the degree to which a change in the linear power
spectrum produces a detectable change in the observable power spectrum
once we allow the galaxy HOD to vary in a way that reproduces $\wpp$. 
A~reduction in $\rms$ to 0.8 can be almost exactly compensated by a change in
the HOD --- in linear theory,
of course, such a change could be exactly compensated by the linear bias 
factor $b_0$.
Linear bias would not change the shape of the observable power spectrum, but
we see in Figure~\ref{fig:dev} that the $\pR(k)$ curves for different $n_s$
and~$h$ converge towards a common shape at $k\ge0.1\hmpci$, though they 
resemble $\plin(k)$ at larger scales. Thus, nonlinear dynamical evolution and
the freedom to introduce scale-dependent bias via HOD variation reduces the
discriminatory power of $\pobs(k)$. However, with the HOD constrained by
$\wpp$, the amplitude of $\pR(k)$ is offset by 5\% or more for 
$0.1\hmpci\leq k\leq0.5\hmpc$ in the $n_s=0.9$ and $h=0.6$ models. The
T06 main galaxy $\pobs(k)$ measurements have typical uncertainty of about
10\% per data point, so this level of offset could provide significant 
additional sensitivity to cosmological parameters if the measurements are
extended to $k=0.5\hmpci$. The caveat is that environmental dependence of
the HOD (not allowed in our models here) might be able to change the
large-scale amplitude of $\pobs(k)$ relative to $\wpp$, erasing the predicted
offset. The systematic uncertainty due to this effect could be removed by
adding and marginalizing over a multiplicative normalization factor
(essentially a ``bias offset'') so that only the shape of $\pobs(k)$ provides
constraints. This technique, and the impact of plausible levels of 
environmental variation, merits further investigation in future work.

\begin{figure}[t]
\centerline{\epsfxsize=3.5truein\epsffile{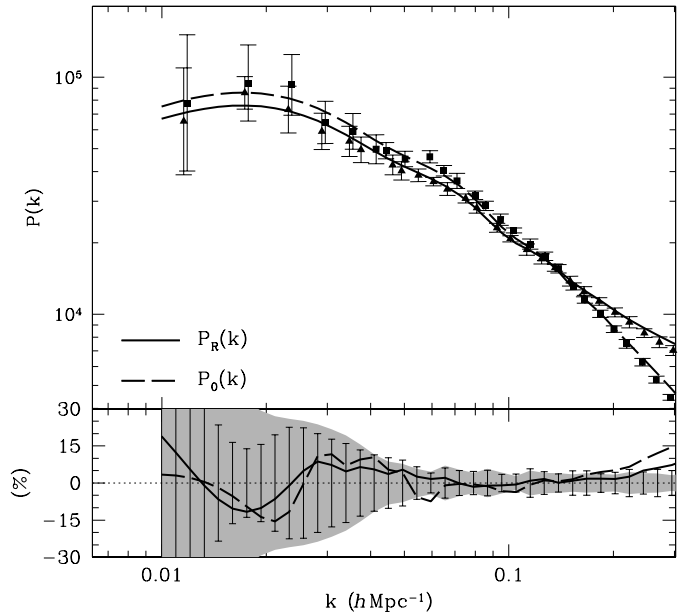}}
\caption{Test of analytic model predictions for $\pR(k)$ and $P_0(k)$, in the 
same format as in Fig.~\ref{fig:lrgtest1}.}
\label{fig:lrgtest2}
\end{figure}

\begin{figure*}[t]
\centerline{\epsfxsize=6.0truein\epsffile{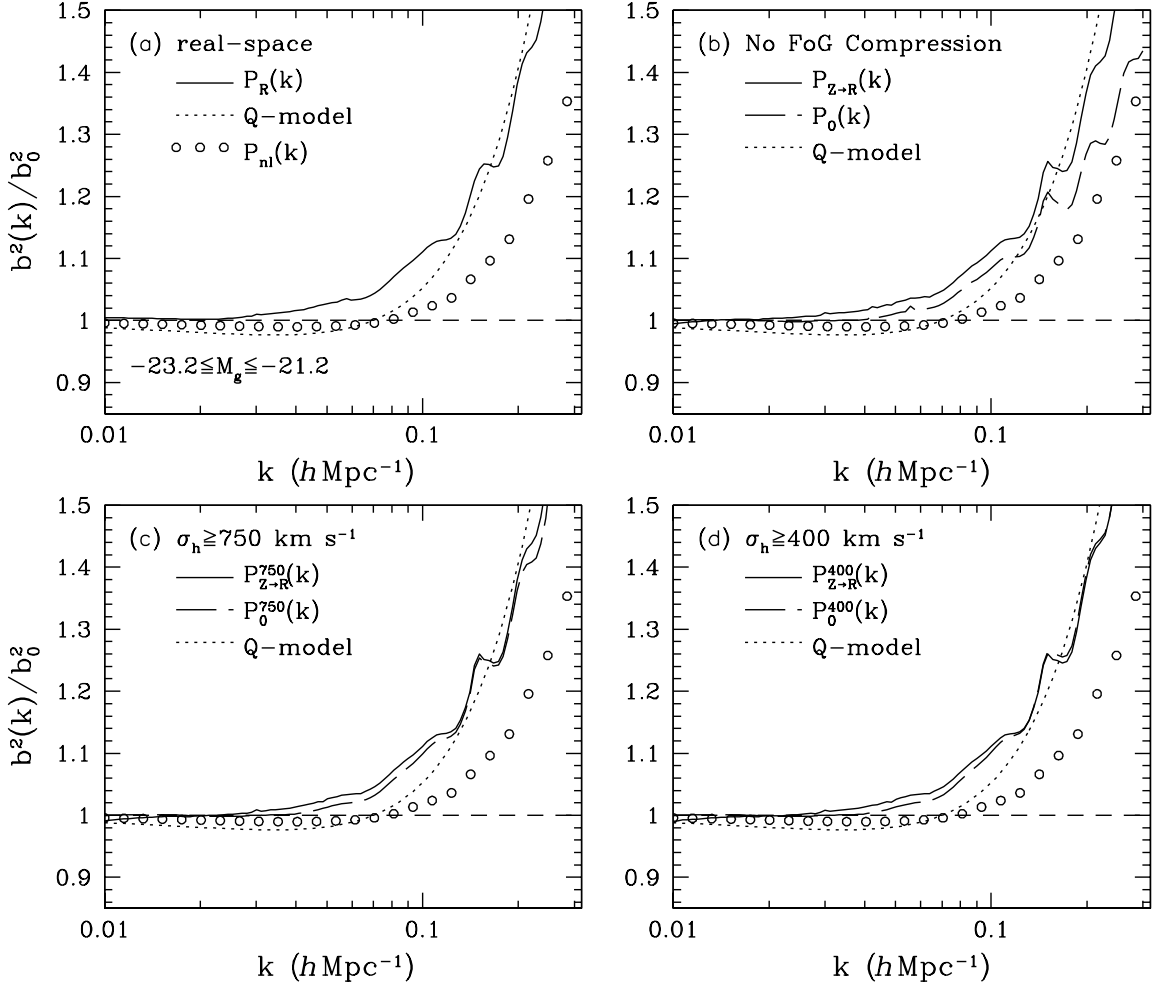}}
\caption{Scale-dependent bias curves for
$\pR(k)$, $\pzr(k)$, and $P_0(k)$ of the luminous red
galaxy (LRG) sample with $-23.2\leq M_g\leq-21.2$, 
in the same format as Fig.~\ref{fig:shape}. Adopting the \citet{zheng5} HOD
parameters, we use the analytic model to predict the scale-dependence of LRG
bias at $z=0.3$. The dotted curve in each panel plots the $Q$-model 
prescription with $Q=20$ that approximately fits our calculations (this
value is somewhat smaller than used in \citealt{max2}, see the text for 
details). Note that $\pnl(k)$ shown as circles is less biased relative to
$\plin(k)$ at $z=0.3$ than at $z=0$.}
\label{fig:lrg}
\end{figure*}

\begin{figure*}[t]
\centerline{\epsfxsize=6.0truein\epsffile{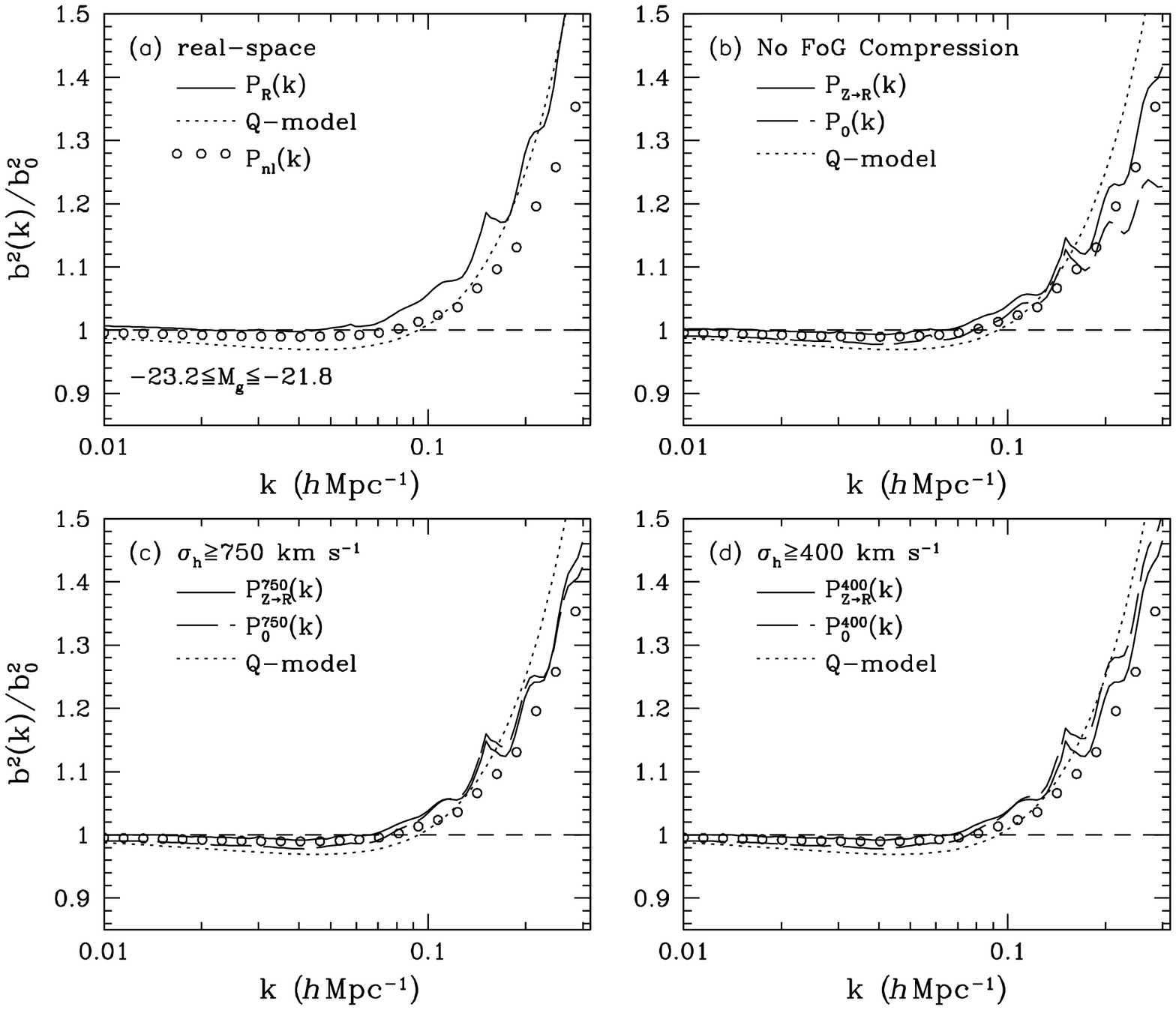}}
\caption{Bias shapes $\pR(k)$, $\pzr(k)$, and $P_0(k)$ of the luminous red
galaxy (LRG) sample with $-23.2\leq M_g\leq-21.8$, 
in the same format as Fig.~\ref{fig:lrg}. The dotted curves are computed
with $Q=15$.}
\label{fig:lrg2}
\end{figure*}

\section{Bias and Power Spectrum of Luminous Red Galaxies}
\label{sec:lrg}
While we have focused on the SDSS main galaxy samples so far, 
the most powerful measurements in galaxy power spectrum
come from the luminous red galaxy (LRG) samples \citep{daniel}
because of the large effective volume that the LRG samples probe.
We consider the bias shape and power spectrum of the LRG samples
separately from those of the main galaxy samples for several reasons.
First, LRGs are physically
distinct from SDSS main galaxies: they are mainly central galaxies, occupying
massive halos of $M\gtrsim10^{13.5}\hmsun$, and they appear to have 
much stronger
scale-dependent bias (e.g., T06 find higher $Q$ values). Second,
constraints on HOD parameters and systematic uncertainties in their values
have not been as extensively investigated; here we draw on parameter
constraints from the recent study of \citet{zheng5}. Third, few simulations 
with the necessary large volume and dynamic range are available for testing our 
analytic model predictions for the LRG samples. The tests of our analytic model
in this regime that we present here
should be regarded as a first step. However, our modeling allows us 
to understand how
the scale-dependence of LRG bias may differ from that of the SDSS main 
galaxies. More thorough investigation of HOD parameters and their 
uncertainties is necessary before applying our method to LRG $\pobs(k)$
measurements to infer cosmological parameters. Here we simply adopt 
HOD parameters from \citet{zheng5} and ignore the impact of uncertainties 
in the parameters for the moment. 

\citet{zheng5} obtain the HOD parameters of the LRG samples with
absolute-magnitude limit $-23.2\leq M_g\leq-21.2$ and
$-23.2\leq M_g\leq-21.8$ by matching
the projected correlation function $\wpp$ and mean space density 
$\bar n_g=9.7\times10^{-5}(\hmpc)^{-3}$ and $2.4\times10^{-5}(\hmpc)^{-3}$,
and by accounting for the
error covariance matrix taken from \citet{idit4} 
(see Appendix~B of \citealt{zheng5} for the HOD parameters).
The mean redshift of the LRG samples is $z=0.3$, and the two LRG samples
are essentially luminosity-thresholded samples. The $M_g\leq-21.2$ sample is
dominated by fainter galaxies, so it probes a smaller volume than the
$M_g\leq-21.8$ sample.

We use a new large $N$-body simulation \citep{warren1} to test our
analytic model predictions for the LRG samples. The simulation is
performed with the Hashed Oct-Tree code \citep{warren2}, evolving 1024$^3$
particles in a volume of comoving $1086\hmpc$ on a side from $z=34$ to the
present, using a $\Lambda$CDM cosmology ($\OM=0.3$, $\Omega_\Lambda=0.7$,
$\Omega_b=0.046$, $h=0.7$, $n_s=1$, and $\rms=0.9$). 
Dark matter halos are identified using the friends-of-friends
algorithm with $b_\up{fof}=0.2$. Since we only have the simulation output at
$z=0$, we cannot directly compute quantities of interest at the mean redshift
of LRGs ($z=0.3$), but we can test whether our analytic model is accurate.

Figure~\ref{fig:lrgtest1} plots the real-space and redshift-space monopole
correlation functions measured from the $N$-body
simulation at $z=0$, populated by
using the same HOD parameters of the LRG sample obtained by fitting $\wpp$ 
measurements at $z=0.3$. The evolution of cosmic structure from $z=0.3$ to
$z=0$ increases the abundance of high mass halos, so with these parameters
we set higher $\bar n_g$ than observed. Note, however,
that we use the simulation just for the purpose of testing our analytic
model. The error bars are computed
by jackknife resampling of the eight octants of the cube.
Our analytic predictions are in good agreement with the populated $N$-body
simulation. The marginally significant discrepancy at $r=4\hmpc$
could indicate that our halo exclusion treatment
is inaccurate at the 10\% level,
but it is difficult to assess the statistical significance of the 
discrepancy with only one simulation. 
Figure~\ref{fig:lrgtest2} shows the analytic predictions for $\pR(k)$ and 
$P_0(k)$, where the statistical uncertainties are computed by using the eight
octants as independent measurements. The analytic model provides good 
approximations to $\pR(k)$ and $P_0(k)$ at $k<0.3\hmpci$.
The shot-noise power spectrum dominates the measurements of $\pR(k)$ and
$P_0(k)$ at $k\gtrsim0.3\hmpci$, and the discrepancy at $k\simeq0.3\hmpci$
may indicate that our shot-noise subtraction scheme is imperfect.
Nevertheless, more simulations are necessary to better quantify the 
statistical significance of the deviation at $k\simeq0.3\hmpci$.
The $N$-body test at $z=0$ demonstrates that our analytic model can be used
to compute
the correlation functions and power spectra in the LRG regime with 
reasonable accuracy, and 
we suspect that these predictions would remain accurate at $z=0.3$.

Figures~\ref{fig:lrg} and \ref{fig:lrg2} plot the bias shapes of the LRG 
samples, computed by using our analytic model and accounting for the mean 
redshift $z=0.3$ of the LRG samples.
We now use the cosmological parameters
$\OM=0.24$, $n_s=0.954$, $h=0.73$, and
$\rms=0.75$, consistent with the WMAP3 results, and LRG HOD parameters for the
same cosmology. The bias shapes are in marked contrast
to those for the $M_r\leq-21$ and $M_r\leq-20$ samples, which would
closely follow
the nonlinear matter power spectrum ({\it circles}).\footnote{We use the
\citet{smith} prescription to compute the nonlinear matter power spectrum.
$\pnl(k)$ at $z=0.3$ is nearly unbiased relative to $\plin(k)$ at 
$k\leq0.1\hmpci$, while at $z=0$ it is biased as is shown in
Figs.~\ref{fig:shape} and~\ref{fig:shape21}. 
In Figs.~\ref{fig:lrg} and~\ref{fig:lrg2}, we plot shorter ranges
along the $x$-axis than in Figs.~\ref{fig:shape} and \ref{fig:shape21},
because of the stronger LRG scale-dependence.}
This is mainly because the fraction of satellite galaxies is small
in the LRG samples, approximately 5\%,
and because their host halos are massive ($>10^{13}\hmsun$) 
and hence highly clustered. 
The bright LRG sample is more biased by 15\% compared to the faint LRG sample,
but it shows less scale-dependence at $k\lesssim0.1\hmpci$.
The low fraction of satellite galaxies also suppresses
the redshift-space multipoles arising from virial motions of satellite galaxies
in halos, and the
$\pzr(k)$ and $P_0(k)$ shapes have little difference compared to the $\pR(k)$
shape. In the bottom panels of Figures~\ref{fig:lrg} and \ref{fig:lrg2},
relatively small changes in the bias shapes arise
between two thresholds of FoG compression, because both cases
basically suppress all the halos that host two or more LRGs. 

The dotted curves plot the $Q$-model prescriptions with $Q=20$ and 
$15~(\hmpc)^2$ for the
faint and bright samples, which
approximately follow our calculations. These values are smaller than the 
best-fit value $Q=30.3~(\hmpc)^2$ found by T06.
Because of the relative statistical weights of the $P(k)$ multipoles, the
quantity $P_{gg}(k)$ that they measure is intermediate between our 
$\pzr(k)$ and $P_0(k)$, but closer to the latter (see the discussion in
\S~\ref{sec:gal} above). The scale-dependence of LRG bias that we predict
from HOD modeling is therefore weaker than that inferred by T06
by fitting the power spectrum. The discrepancy may reflect the difference 
between the ``defogging'' procedure used by T06 and the perfect FoG
compression assumed here, though since we find that FoG compression has little
impact this explanation would imply that the T06 method overcorrects
FoGs. (More direct evidence for such overcorrection is presented by B.~Reid
et~al. [in preparation]). It should also be emphasized that
$no$ values of $Q$ can fit our calculations with 5\% accuracy
up to $k=0.2\hmpci$. However, we do confirm a basic result of T06
model fitting: scale-dependence of LRG bias becomes substantial at 
$k>0.1\hmpci$, and it is much stronger than that of less luminous, main
sample galaxies.

\begin{figure}[t]
\centerline{\epsfxsize=3.5truein\epsffile{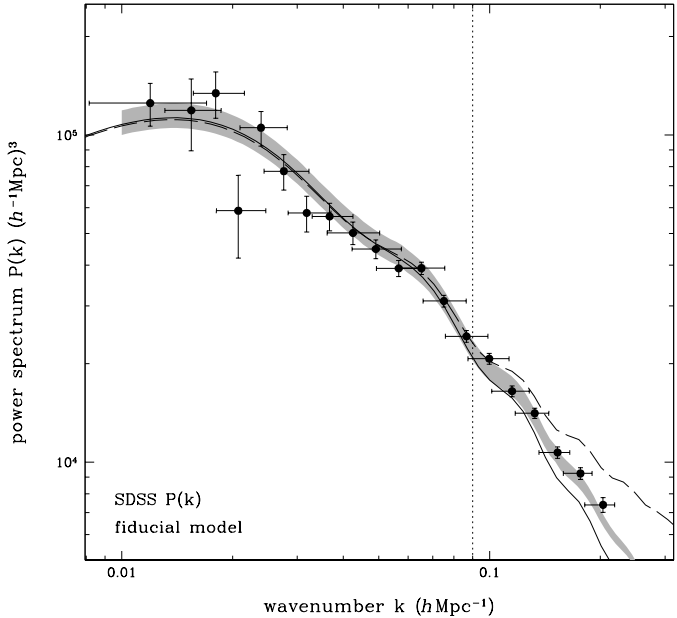}}
\caption{Predicted LRG power spectrum and the SDSS measurements.
Points are the power spectrum measurements of the flux-limited LRG sample 
without defogging, taken from \citet{max2} (see the text for the difference
in the analyzed samples). Using our analytic model, we plot 
$\hat P_{gg}(k)\equiv0.8P_0(k)-0.07P_2(k)+0.006P_4(k)$, 
which closely matches the power spectrum estimate used in \citet{max2},
after statistical weights on multipole measurements are considered.
The shaded gray bands show our
predicted power spectrum of the volume-limited LRG samples with the lower bound 
for the LRG sample with $-23.2\leq M_g\leq-21.2$ and the upper
bound for the LRG sample with $-23.2\leq M_g\leq-21.8$. 
The solid line 
represents $\plin(k)$ of our fiducial cosmological model and the dashed line is
the $Q$-model prediction with $Q=30.3$ quoted in \citet{max2} that best fits
their ``defogged'' measurements. 
The vertical dotted lines represent the ``nominal'' nonlinear scale quoted
by \citet{max2}, beyond which the measurements provided little leverage on
cosmological parameters.}
\label{fig:lrg-pow}
\end{figure}

\begin{figure*}[t]
\centerline{\epsfxsize=6.0truein\epsffile{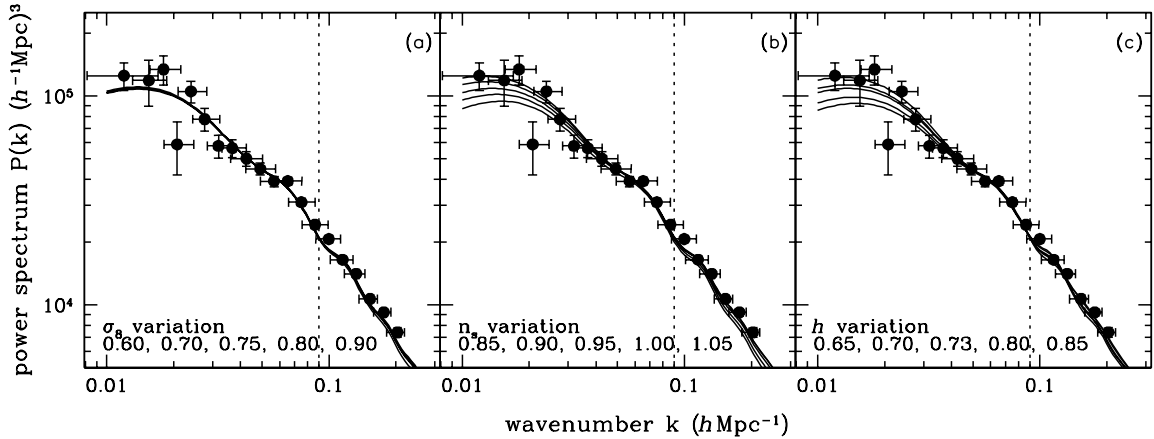}}
\caption{Effect of cosmological parameter variations on the
predicted LRG power spectrum, in the same format as Fig.~\ref{fig:lrg-pow}.
For each variation of our adopted fiducial cosmological
model shown in the legend, we compute power spectra of the two volume-limited
LRG samples and solid lines are
the average power spectra weighted by their number density.
The sequence of $\hat P_{gg}(k)$ with varying $\rms$ in Panel~$(a)$ is nearly
identical, while the predictions of $\hat P_{gg}(k)$ in Panels~$(b)$ and~$(c)$
decrease at $k=0.01\hmpci$ with increasing $n_s$ or $h$.}
\label{fig:lrg-pow2}
\end{figure*}

\begin{figure*}[t]
\centerline{\epsfxsize=6truein\epsffile{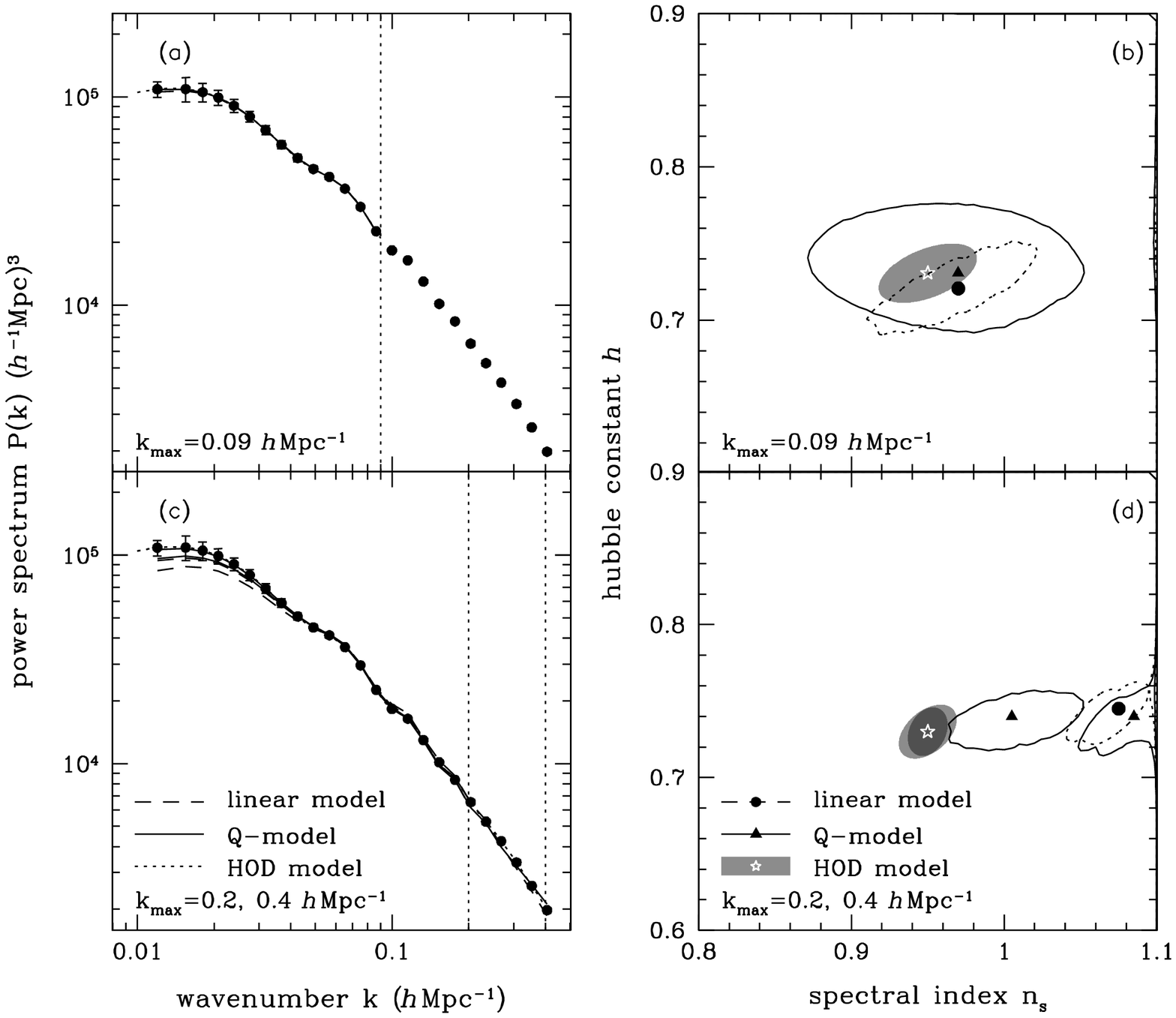}}
\caption{Constraints on cosmological parameters obtainable from a future 
``final SDSS'' LRG sample, 
using linear theory, $Q$-model, and HOD model fitting.
Left panels show the synthetic data with error bars and best-fit
power spectra for the three models as smaller 
scale measurements are included in
the data fitting. Right panels plot the 1-$\sigma$ confidence levels 
($\Delta\chi^2=2.3$) and 
best-fit cosmological parameters ({\it symbol}) inferred by
fitting the three models to the synthetic data.}
\label{fig:contour}
\end{figure*}

Proper application of our method to observations will require several key steps
beyond the scope of our current investigation. First and foremost is 
careful matching of galaxy samples analyzed for $\pobs(k)$ measurements and 
HOD modeling. T06 use ``flux-limited'' LRG and main 
galaxy catalogs from SDSS DR4 to obtain $\pobs(k)$ measurements
and constraints on the scale-dependent bias, while we use Zheng et~al.'s (2008)
``volume-limited''
galaxy samples to constrain HOD parameters
and predict $\pobs(k)$. Second, as mentioned earlier, we need more large-volume
simulations at $z>0$ 
for testing the accuracy of our analytic model in a regime adequate for
LRG clustering. Third, we need to more fully investigate HOD uncertainties
associated with analytic model fitting 
of $\wpp$ and with possible environmental dependence of the LRG HOD.

With these caveats in mind, we present preliminary results on comparison of
our analytic model predictions to the
T06 measurements of the LRG power spectrum
in Figure~\ref{fig:lrg-pow}. Since it is hard to correctly account for the
difference in our FoG compression and their defogging processes, we instead
use the $\pobs(k)$ measurements without defogging (see Fig.~22 in 
T06 for the no-defog data).
Points with error bars show the SDSS measurements.

In Figure~\ref{fig:lrg-pow}, the 
shaded gray bands show $\hat P_{gg}(k)\simeq0.8P_0(k)-0.07P_2(k)+0.006P_4(k)$
for the two LRG samples,
constructed by using our analytic model predictions of $P_0(k)$, $P_2(k)$,
and $P_4(k)$, with the lower bound computed
from the faint sample and the upper bound from the bright sample
(see the Appendix of T06 for the empirical formula for 
$\hat P_{gg}(k)$).
Considering the difference in the analyzed galaxy samples, the
shaded region is likely to encompass the true $\pobs(k)$, and it is indeed 
in good agreement with the measurements. Note that our analytic model 
prediction allows no freedom in the asymptotic bias $b_0$, and hence in the
normalization of $\hat P_{gg}(k)$, once HOD parameters
are pinned down by fitting $\wpp$.\footnote{Of 
course, it is no surprise that power spectrum and $\wpp$ measurements imply
similar amplitudes of galaxy clustering, but this agreement is nonetheless
reassuring.} This  is in contrast to the linear or $Q$-model
fit, in which $b_0$ can be arbitrarily adjusted to fit the data.
The solid and dashed lines show $b^2_0\plin(k)$ 
of the fiducial cosmological model and the $Q$-model prescription
with $Q=30.3~(\hmpc)^2$ quoted in T06, which 
best fits their ``defogged''
measurements.\footnote{Since T06 provide best-fit $Q$-values for
defogged data only, the dashed line should be considered as a reference,
rather than a comparison. Note that their defogged data would roughly follow 
the dashed line.}
At $k\leq0.09\hmpci$, all predictions are consistent with the linear
model ({\it solid}) at roughly
the 10\% level, justifying the applicability of the linear model with the
same accuracy. However, significant scale-dependence of LRG bias prevents
the use of the linear model at $k\geq0.1\hmpci$. The vertical dotted lines
show the nominal nonlinear scale quoted by T06; they fit $Q$-models
up to $k=0.2\hmpci$ but note that data beyond $k=0.1\hmpci$ provided little
leverage on cosmological parameters and mainly constrained $Q$.

We further investigate this point by
considering variations of the fiducial cosmological parameters in 
Figure~\ref{fig:lrg-pow2}.
For each cosmological model variation, we first re-fit
$\wpp$ to obtain best-fit HOD parameters and then compute $\hat P_{gg}(k)$.
The solid lines are computed by averaging $\hat P_{gg}(k)$ of the two LRG
samples with weight by their number density.
Figure~\ref{fig:lrg-pow2}$a$ 
shows a sequence of $\hat P_{gg}(k)$ with varying $\rms$ values. 
The predictions are virtually identical, showing that the combination of
$\hat P_{gg}(k)$ and $\wpp$ has no direct constraining power on $\rms$
given the flexibility of our 5-parameter HOD model. Since the predictions are 
obtained by applying our analytic model of $\hat P_{gg}(k)$ to Zheng
et~al.'s (2008) HOD fits to $\wpp$, both of which depend on $\rms$ in a
complex way, this constancy of $\hat P_{gg}(k)$ is a reassuring consistency
check.

In Figures~\ref{fig:lrg-pow2}$b$ and~\ref{fig:lrg-pow2}$c$, we 
consider sequences with varying $n_s$ and $h$, respectively. 
Once we have fit $\wpp$ for a specific cosmology, there is no freedom to
adjust the shape or amplitude of the model $\hat P_{gg}(k)$, except within
the observational uncertainties on the HOD parameters. We see from 
Figures~\ref{fig:lrg-pow2}$b$ and~\ref{fig:lrg-pow2}$c$
that normalizing to $\wpp$ effectively forces
different models to agree at $k\simeq0.08\hmpci$. The largest separation among
the models comes at large scales, $k\lesssim0.02\hmpci$. However, the models 
also
differentiate noticeably on smaller scales, especially when $n_s$ is varied.
In linear theory fits, these scales would be discarded because the model
predictions are unreliable. In $Q$-model fits, the information they contain
would largely be lost by marginalizing over $Q$. The varying $n_s$ models also
predict different $\wpp$ for $r_p>20\hmpc$, but these scales carry little
weight in the HOD fitting.

These results suggest that fitting $\pobs(k)$ to small scales with HOD 
parameter constraints from $\wpp$ could yield significantly improved constraints
on cosmological parameters. To investigate this point, and to compare linear
theory, $Q$-model, and full HOD-model fitting, we generate synthetic 
$\pobs(k)$ measurements equal to the predictions of our analytic model for the
fiducial cosmological parameters and the corresponding \citet{zheng5} 
HOD parameters. We assign observational errors that are a factor of two
smaller than those found by T06 at the same value of $k$, roughly
approximating the improvement that might come from the future full SDSS data
set.\footnote{Error bars from the Baryonic Oscillation Spectroscopic Survey
(BOSS) of SDSS-III will be much smaller still.}
At $k>0.2\hmpci$ (where T06 did not present measurements), we assign
fractional uncertainties equal to those at $k=0.2\hmpci$.

Figure~\ref{fig:contour} shows the synthetic measurement and its
errors with our analytic model prediction ({\it dotted}) in the left panels
and the cosmological parameter constraints inferred by fitting these data
using the three different models in the right panels.
For simplicity, we fixed the combinations
$\OM h^2=0.1272$ and $\OB h^2=0.0222$, which
are best constrained from cosmic microwave background measurements,
and we varied $n_s$ and $h$ as two free parameters, assuming
a flat $\Lambda$CDM universe. 
In Figure~\ref{fig:contour}$a$, the solid and dashed lines
show the best-fit linear model and $Q$-model using the 
measurements only at $k\leq k_\up{max}=0.09\hmpci$.
We marginalize over $b_0$ for linear theory and $(b_0,Q)$ for 
$Q$-model fitting. Both models can describe the data reasonably well 
(the dashed line is largely obscured by the solid line).
The contours and shaded ellipses 
in Figure~\ref{fig:contour}$b$ show the 1-$\sigma$
confidence levels ($\Delta\chi^2=2.3$)
for all three models, and the symbols represent the best-fit 
parameters. 
To compute the parameter constraints of the HOD model, we use the Fisher 
matrix formalism and marginalize over
a cosmological parameter $\rms$ and two sets of five HOD parameters
for each LRG sample. Note that analytic model fitting of $\wpp$
assumes a $\rms$ value, which is largely degenerate in $\pobs(k)$ measurements.

The HOD model ({\it shaded region})
provides tighter constraints than linear theory
({\it dashed contour}), because we have additional information from $\wpp$
measurements. The best-fit parameters ({\it circle}) of the linear theory fit
are only marginally consistent at 1-$\sigma$ level with the true values 
({\it asterisk}),
because the linear bias approximation is no longer an
accurate description of LRG bias
even at $k\leq0.09\hmpci$, given the small uncertainties in $\pobs(k)$
assumed here.
The $Q$-model prescription ({\it solid contour})
yields relatively less biased best-fit parameters ({\it triangle}).
However, marginalizing over $Q$ compromises
statistical constraining power, and this fit has the largest 
uncertainties in parameter estimates among the three models.

The bottom panels compare the three models with $k_\up{max}=0.2$ and
$0.4\hmpci$, demonstrating the utility of higher $k$ measurements.
The best-fit $\pobs(k)$ ({\it dashed line}) of linear theory 
in Figure~\ref{fig:contour}$c$ shows large deviations, and
the best-fit parameters ({\it circles}) in Figure~\ref{fig:contour}$d$
are significantly different from the true values 
(out of range $n_s>1.1$ for $k_\up{max}=0.4\hmpci$). The linear theory fit
prefers higher $n_s$ to compensate the deficit in $\pobs(k)$ 
on small scales, while the overall $\pobs(k)$ shape and baryon wiggles 
keep the
$h$~value relatively unchanged. Clearly, linear theory is an invalid
description of $\pobs(k)$ for LRGs
when $k_\up{max}>0.09\hmpci$.

A similar but less extreme
trend of bias in best-fit parameters is found when the $Q$-model
prescription is applied to $\pobs(k)$ with $k_\up{max}\geq0.2\hmpci$.
The best-fit parameters of the $Q$-model fit are (1.0, 0.74), deviating from 
the true value and from the
best-fit values with $k_\up{max}=0.09\hmpci$ at the 1-$\sigma$ level. 
Higher $n_s$ is again favored, but a higher~$Q$ 
can partially balance the trend of higher $n_s$. 
With $k_\up{max}=0.4\hmpci$, the best-fit parameters are (1.09, 0.74) and
biased at the 3-$\sigma$ level. The $Q$-model prescription also develops 
substantial bias in best-fit parameters as $k_\up{max}$ increase. In contrast,
while our HOD prediction is accurate by construction in this experiment, 
it makes full use of the variations of $\pobs(k)$ on small 
scales, providing tighter constraints on cosmological parameters
as small scale measurements are included.

Since we have generated the synthetic data from the HOD model, we have adopted
the most optimistic scenario in which the HOD model with correct parameters
can give a perfect description of the data. Nevertheless, 
Figure~\ref{fig:contour} conveys two key points. First, $Q$-model fitting to
the LRG $\pobs(k)$ may yield unreliable parameter estimates for 
$k_\up{max}=0.2\hmpci$ (and linear theory may be problematic even at
$k_\up{max}=0.09\hmpci$). Second, increasing $k_\up{max}$ from $0.09\hmpci$
to $0.2\hmpci$ with HOD model fitting can significantly reduce statistical
uncertainties in cosmological parameters, and increasing to $0.4\hmpci$ 
produces a modest further gain.

\section{Conclusions}
\label{sec:sum}
We have developed an analytic model to predict observable galaxy power spectra
$\pobs(k)$ for specified cosmological and galaxy HOD parameters, and 
we have verified its accuracy using $N$-body simulations.
As potentially observable power spectra $\pobs(k)$, 
we have considered the real-space $\pR(k)$, the redshift-space
monopole $P_0(k)$, and the pseudo real-space $\pzr(k)$, with varying levels of
Finger-of-God (FoG) compression for the latter two. Once HOD parameters are 
determined by fitting the number density $\bar n_g$ and projected correlation 
function $\wpp$ of the observed SDSS galaxy samples, given a specified 
cosmological model, our analytic model can be used to predict $\pobs(k)$
measurements. 

The large-scale normalization of our predictions is also fixed
in the process of fitting $\wpp$, providing a unique prediction for
each combination of cosmological and HOD parameters.
In practice, one can simultaneously fit cosmological and HOD
parameters using $\pobs(k)$ and $\wpp$ as constraints, then marginalize over
the HOD in deriving cosmological parameters. By implementing a complete
physical model of nonlinear galaxy bias and drawing on the additional
information in $\wpp$, our method allows one to take full advantage
of precision measurements of $\pobs(k)$ on quasi-linear scales 
($k=0.1-0.4\hmpci$), where linear theory or the phenomenological $Q$-model
may be insufficiently accurate. Our main findings are as follows:

1. Our analytic model for calculating $\wpp$ follows the method described in
\citet{jeremy}, with the improved treatment of the scale-dependent halo bias
and ellipsoidal halo exclusion corrections. Drawing on the \citet{jeremy2}
model for redshift-space distortion, the analytic model is extended to
incorporate calculating real-space and redshift-space power spectra.
We have tested its predictions for $\wpp$ and $\pobs(k)$ against populated 
$N$-body simulations spanning cosmological parameter range
$\OM=0.1-0.63$ and $\rms=0.6-0.95$, with HOD parameters matched to represent 
two SDSS galaxy samples with absolute-flux limits $M_r\leq-20$ and $M_r\leq-21$
\citep{idit3}. The analytic model reproduces the numerical results of $\wpp$
to 5\% or better, and the
predictions of $\pobs(k)$ are consistent with the numerical results to 2\%
at $k=0.1-1\hmpci$ and to 10\% at $k=0.025-0.1\hmpci$, though the finite box
size of the simulations makes it difficult to assess the statistical 
significance of differences on large scales.

2. For the $M_r\leq-20$ galaxy sample, the pseudo real-space power spectrum
$\pzr(k)$ recovers the true $\pR(k)$ to 2\% at $k\leq0.2\hmpci$, while the 
deviation between $\pR(k)$ and the scaled monopole $P_0(k)$ is already 10\% at 
$k=0.1\hmpci$. However, the deviation of $\pzr(k)$ from $\pR(k)$ becomes
substantial at $k\geq0.3\hmpci$. This deviation can be partly remedied by FoG 
compression, which suppresses nonlinear behavior of the redshift-space
multipoles caused by the random motions of satellite galaxies within halos.
With FoG compression threshold $\sigma_\up{h}=750~\kms$,
$\pzrF{750}(k)$ can recover $\pR(k)$ to 5\% at $k\leq0.45\hmpci$,
and at higher $k$ for $\pzrF{400}(k)$. FoG compression also
reduces nonlinearity of the monopole power spectrum, but $P_0^{400}(k)$ can only
achieve 10\% accuracy at $k\leq0.3\hmpci$. We conclude that the pseudo
real-space method of \citet{max0} is an effective tool for recovering the
nonlinear real-space galaxy power spectrum from redshift-space measurements,
especially if it is combined with accurate FoG compression.

3. The nonlinear $matter$ power spectrum describes the nonlinear real-space 
galaxy power spectra to 1\% at $k\leq0.2\hmpci$ for the $M_r\leq-20$ and 
$M_r\leq-21$ galaxy samples, up to an overall bias factor $b^2_0$. The shape 
of the scale-dependent bias function $b^2(k)/b_0^2$ for $\pzr(k)$ is 
qualitatively similar to
$\pR(k)$ at $k\leq0.3\hmpci$, but the shape for $P_0(k)$ is completely
different over the entire range we consider here. FoG compression makes little
difference to $b^2(k)/b^2_0$ for $\pzr(k)$, but a large difference for $P_0(k)$.
For these SDSS main galaxy samples, the $Q$-model prescription
traces our calculation of $\pR(k)$ relatively well at $k\geq0.1\hmpci$, but
its shape on large scales differs, so it might induce some overall bias 
in cosmological parameters when fitted to
$\pobs(k)$ measurements that have large uncertainties at $k\leq0.05\hmpci$.
Similar trends but with larger discrepancy are found in comparison to 
our $\pzr(k)$ and $P_0(k)$ calculations.

4. Uncertainties in computing $\pobs(k)$ in our method arise from observational
uncertainties in the HOD parameters and from uncertainty in the adopted
parametrization itself.
We have examined these uncertainties by adopting a flexible 
HOD parametrization with freedom to explore a wider range of plausible halo 
occupation functions. For the $M_r\leq-20$ sample with the \citet{idit3}
uncertainties in $\wpp$, the uncertainty in the predicted $\pobs(k)$
is 2\% at $k=0.2\hmpci$, becomes progressively smaller
at lower $k$, and climbs up to 4\% at $k=0.5\hmpci$. The uncertainty is
a factor of two smaller for the $M_r\leq-21$
sample, roughly the ratio of the fractional $\wpp$ measurement errors
of the two samples. We have not investigated the uncertainties associated
with possible environmental variations of the HOD \citep{croton}. Based on
work to date, we expect that such variations might lead to few percent
uncertainties in the overall normalization predicted for $\pobs(k)$ after
fitting $\wpp$, but that the impact on scale-dependence of $b^2(k)/b^2_0$
would be smaller.

5. Moving to the LRG regime, we have tested our analytic model predictions
against the $z=0$ output of a large volume, 1024$^3$-particle $N$-body 
simulation \citep{warren1}, populated based on Zheng et~al.'s (2008) HOD
fits to $\wpp$ for two volume limited SDSS LRG samples \citep{idit4}.
The analytic model predicts $\xi_0(r)$ and $\xiR(r)$ to 5\% or better over 
the range $0.1\hmpc\leq r\leq30\hmpc$, and the predictions for $P_0(k)$ and 
$\pR(k)$ have similar accuracy over the range $0.01\hmpci\leq k\leq0.3\hmpci$.

6. For the LRG samples, the linear (scale-independent) bias approximation 
remains accurate at the 5\% level to $k=0.08\hmpci$ for the
 $-23.2\leq M_g\leq-21.2$ sample and to $k=0.1\hmpci$ for the
 $-23.2\leq M_g\leq-21.8$ sample. There is little variation among
$\pzr(k)$, $P_0(k)$, and $\pR(k)$, because LRGs are mainly central
galaxies in massive halos, so random motions of satellite galaxies have little
impact. Similarly, FoG compression has only a small impact on $b^2(k)/b_0^2$
for these samples. Both samples show strong scale-dependence of bias 
at $k\geq0.1\hmpci$, much more than for main sample galaxies.\footnote{This
result, obtained by fitting HODs and computing $\pobs(k)$ with our analytic
model, confirms the result of T06 inferred by fitting $\pobs(k)$ with
$Q$-models. However, the actual scale-dependence we find for LRGs is somewhat
weaker than that inferred by T06.} If we fit $b^2(k)/b_0^2$ from our
HOD models with the best $Q$-model over the range $k=0.01-0.2\hmpci$, the 
largest deviation is 7\%.

7. We have presented a preliminary comparison of our analytic model 
predictions to the T06 measurements of the LRG $\pobs(k)$, with no
FoG compression. The difference between our volume-limited samples and the
T06 flux-limited sample precludes a full quantitative assessment, but the
qualitative agreement is remarkably good over the full range of the 
measurements, $k=0.01-0.2\hmpci$ (Fig.~\ref{fig:lrg-pow}). Fits with different
cosmological parameters differ on large scales and, to a smaller degree,
at $k\geq0.2\hmpci$, indicating that measurements to smaller scales would
provide additional discriminatory power.

8. Looking to the future, we have generated synthetic $\pobs(k)$ data from
our analytic model with error bars half those of T06, then fit them to
successively higher $k_\up{max}$ with linear theory, the $Q$-model, and the
HOD model. Cosmological parameters from linear theory fits are badly biased
for $k_\up{max}\geq0.1\hmpci$, while for $k_\up{max}=0.09\hmpci$ they are
biased at less than 1-$\sigma$. Parameters from the $Q$-model are minimally
biased for $k_\up{max}=0.09\hmpci$, biased by 1.2-$\sigma$ for 
$k_\up{max}=0.2\hmpci$, and biased by many-$\sigma$ for $k_\up{max}=0.4\hmpci$.
Since the synthetic data are generated from the HOD model, the HOD parameter
estimates are unbiased, and the error bars in cosmological parameters shrink
steadily as $k_\up{max}$ is increased from $0.1\hmpci$ to $0.2\hmpci$ to
$0.4\hmpci$. 

Results~7 and~8 are especially encouraging. Using only the HOD model and the
information in $\wpp$, our method predicts exactly the scale-dependent bias
for LRGs that is required to transform the linear power spectrum from WMAP3
into the SDSS galaxy power spectrum measured by T06. This is in contrast to
$Q$-model fitting, where a phenomenological parameter (motivated by simulation
results but with no clear physical interpretation) is introduced specifically 
to account for the difference between the linear theory $P(k)$ and the
observed power spectrum.

Despite the clear evidence that scale-dependent bias affects the LRG power
spectrum beyond $k=0.1\hmpci$, result~8 shows that one can gain substantial
additional leverage on cosmological parameters with HOD modeling of power 
spectrum measurements up to $k=0.2-0.4\hmpci$, and possibly beyond. Realizing
this opportunity will require several investigations beyond those presented
here. First, we will need more large volume simulations to test and, if
necessary, refine our analytic model to the level of accuracy demanded
by the final SDSS data set. Second, we must explore more thoroughly the
uncertainties associated with the HOD fitting, including alternative 
parametrizations, the impact of velocity bias on redshift-space predictions,
and the possible impact of environmental variations of the HOD.
Given the growth of current and future galaxy surveys in depth and redshift,
these investigations will be needed to go beyond linear theory.
Precise measurement of the primordial matter power spectrum will play
a crucial role in constraining cosmological parameters and testing dark
energy theories.

\acknowledgments
We are grateful to Max Tegmark for kindly providing his no-defog measurements 
used in our Figure~\ref{fig:lrg-pow}.
J.~Y. has been supported by the Harvard College Observatory through a 
Donald~H. Menzel Fellowship and by the Graduate School of The Ohio State 
University through a Presidential Fellowship. D.~W. acknowledges support from
NSF Grant AST-0407125. J.~T. was
supported by the Chandra award GO5-6120B and National Science
Foundation (NSF) under grant AST-0239759.
Z.~Z. gratefully acknowledges support from the Institute for Advanced
Study through a John Bahcall Fellowship.

\end{document}